\documentclass[prb,twocolumn,superscriptaddress,amsmath,amssymb,floatfix]{revtex4-2}

\usepackage{graphicx}
\usepackage{booktabs}
\usepackage{bm}
\usepackage[colorlinks=true,linkcolor=blue,citecolor=blue,urlcolor=blue]{hyperref}

\graphicspath{{figures/}{./}}


\newcommand{\Kp}[1]{\mathcal{K}^{(#1)}}
\newcommand{\eps}{\epsilon_M}
\newcommand{\Pexc}{P_{\rm exc}}
\newcommand{\Pbar}{\bar P_{\rm exc}}
\newcommand{\Pleak}{P_{\rm leak}}
\newcommand{\Pflip}{P_{\rm flip}}
\newcommand{\tret}{\tau_{\rm return}}
\newcommand{\Gesc}{\Gamma_{\rm esc}}
\newcommand{\evum}{eV\,$\mu$m$^{-3}$}
\newcommand{\effu}{eV$^{-1}\,\mu$m$^{3}$}

\begin{document}

\title{Parity-Resolved Many-Body Kinetics of a Topological Qubit under Nonequilibrium Phonon Pulses}

\author{Han Huang}
\email{hh697@cornell.edu}
\affiliation{Sibley School of Mechanical and Aerospace Engineering, Cornell University, Ithaca, New York 14853, USA}

\date{\today}

\begin{abstract}
Phonon bursts from ionizing radiation cause correlated errors in superconducting quantum processors. We study prescribed nonequilibrium phonon pulses in a Majorana qubit formed by two $2\,\mu\mathrm{m}$ proximitized nanowires. Combining the lattice Lutchyn--Oreg model, microscopic electron--phonon coupling, and parity-resolved many-body dynamics, we track excitation, relaxation, and logical coherence. Without quasiparticle escape, recombination restores the initial Majorana occupations in about 130 ns, while phonons exchanged during excitation and recovery can retain logical-state information, leaving permanent coherence loss. For mirror-symmetric wires, the fractional coherence loss per wire is 0.75--1.0 times the excitation probability averaged over the two logical states, even at zero Majorana splitting; breaking mirror symmetry can suppress dephasing at small splitting. Factorized Boltzmann kinetics neglects quasiparticle correlations and overestimates residual parity flips. At equal deposited energy density, phonon spectra yield different excitation probabilities and coherence losses. Comparisons with Kitaev chains matched to the wire's gap and Majorana decay length reveal that low electron density suppresses excitation by energetic phonons, whereas hard-wall boundaries in a high-density control increase excitation efficiency by a factor of $5\times10^4$ relative to the wire for pulses centered at 1 meV. Magnetic-field and chemical-potential tuning yield different phonon sensitivities at the same bulk gap through changes in spin and electron--hole structure. Including hybrid states' semiconductor weights reduces near-threshold excitation rates fourfold when the tunnel coupling equals the parent superconducting gap. Our framework connects phonon spectra to leakage, poisoning, and coherence loss to guide Majorana-device design.

\end{abstract}

\maketitle

\section{Introduction}

Topological qubits encode quantum information in the fermion parities of spatially separated Majorana zero modes. In the topological regime, these modes can form at the ends of semiconductor nanowires coupled to a conventional superconductor~\cite{Kitaev2001,Lutchyn2010,Oreg2010,Alicea2012,Lutchyn2018,Aghaee2023,Aghaee2025}. Quasiparticle exchange with the environment changes a wire’s parity and causes poisoning, which topological protection alone does not prevent~\cite{Rainis2012,Karzig2021}. Ionizing radiation provides a possible source of quasiparticle excitations by generating nonequilibrium phonons that propagate through the substrate. In conventional superconducting circuits, these phonons can break Cooper pairs and cause correlated errors across multiple qubits~\cite{Wilen2021,McEwen2022,Martinis2021,Huang2024}. They can also excite bulk quasiparticles in proximitized nanowires. Understanding their effect on Majorana qubits requires answering three questions: which phonons excite the wires, whether the Majorana occupations recover after the pulse, and whether the stored quantum superposition survives.

Previous studies address parts of this problem using different descriptions of the electronic states and their environment. Phonon effects in Majorana wires have been studied for thermal baths~\cite{Aseev2019,Song2018,Knapp2018,Kaestle2020}, while radiation models for conventional superconducting circuits focus on phonons above the pair-breaking threshold of aluminum~\cite{Martinis2021,Yelton2024}. A nonequilibrium burst has a broader spectral structure, and proximitized wires can also absorb phonons below that threshold. Mean-density and distribution-function approaches~\cite{Rothwarf1967,Kaplan1976,Huang2024,Karzig2021} describe recombination when quasiparticles encounter one another independently~\cite{WilsonProber2004}, but can miss the correlations within the few pairs created by a weak pulse. Reservoir-arrival models instead describe quasiparticles entering a wire and changing its parity~\cite{Rainis2012}. Master equations have treated quasiparticle counts in small islands~\cite{Maisi2013}, Andreev levels coupled to a thermal continuum~\cite{Olivares2014}, and Kitaev chains in thermal baths~\cite{PedrocchiDiVincenzo2015,Pedrocchi2015,Ippoliti2016,Breckwoldt2022}, without directly connecting nonequilibrium phonon spectra to correlated, mode-resolved dynamics in semiconductor wires. Coherence has also been studied under excitation~\cite{Schmidt2012,Cheng2012,Alase2025}, prescribed error probabilities~\cite{Knapp2018Quantum}, classical noise~\cite{Mazza2013,Mishmash2020,Konschelle2013}, and thermal baths with model couplings~\cite{Ippoliti2016,Munk2019}. For a microscopic phonon pulse, however, it remains unanswered how excitation and relaxation jointly affect populations and coherence, including whether exchanged phonons retain information about the logical state~\cite{Akhmerov2010}.

In this work, we calculate the response of a four-Majorana qubit formed by two proximitized nanowires to prescribed nonequilibrium phonon pulses. We obtain the discrete quasiparticle states from the lattice Lutchyn--Oreg model~\cite{Lutchyn2010,Oreg2010} and calculate transition rates from microscopic deformation-potential and piezoelectric couplings. These rates resolve the phonon energy, polarization, and propagation direction. We evolve the joint occupation probabilities of the retained modes using a parity-conserving Pauli master equation that preserves pair correlations, and compare its predictions with factorized Boltzmann kinetics, which treats quasiparticle occupations as statistically independent and therefore neglects these correlations. We then evolve the logical coherence using the many-body density matrix to determine whether the stored quantum superposition survives excitation and recombination, accounting for information from the logical state carried away by phonons. To quantify the correction from semiconductor--superconductor hybridization~\cite{Stanescu2017}, we compare excitation rates calculated with and without the semiconductor weight of hybrid quasiparticle states. To test which results require the semiconductor band structure, we compare the nanowire with spinless Kitaev chains~\cite{Kitaev2001} having the same gap and Majorana decay length, at both comparable and higher electron densities.

For $2~\mu\mathrm{m}$ wires with InAs-like effective parameters, recombination restores the initial Majorana occupations on a mean timescale of about 130 ns when quasiparticles cannot escape. Permanent parity poisoning requires quasiparticle exchange with the environment, whereas factorized Boltzmann kinetics predicts a spurious residual Majorana occupation of roughly twice the excitation probability in the weak-pulse limit. Parity recovery nevertheless leaves a loss of logical coherence because the exchanged phonons can distinguish the logical states. At equal deposited energy density, different pulse spectra produce substantially different excitation probabilities and coherence losses. The electronic structure further controls this response: at low electron density, momentum mismatch suppresses excitation by high-energy phonons. At zero chemical potential, a selection rule also weakens the coupling of phonons to the low-energy states near the field-driven transition. Consequently, reducing the gap by tuning the magnetic field does not strongly enhance phonon excitation. These results show why neither the excitation gap nor parity recovery alone determines resilience to phonon bursts. Our framework provides a microscopic basis for assessing radiation sensitivity and guiding Majorana-device design.

\section{Model}\label{sec:model}

\subsection{Qubit and wire spectrum}

\begin{figure*}[t]
\includegraphics[width=\textwidth]{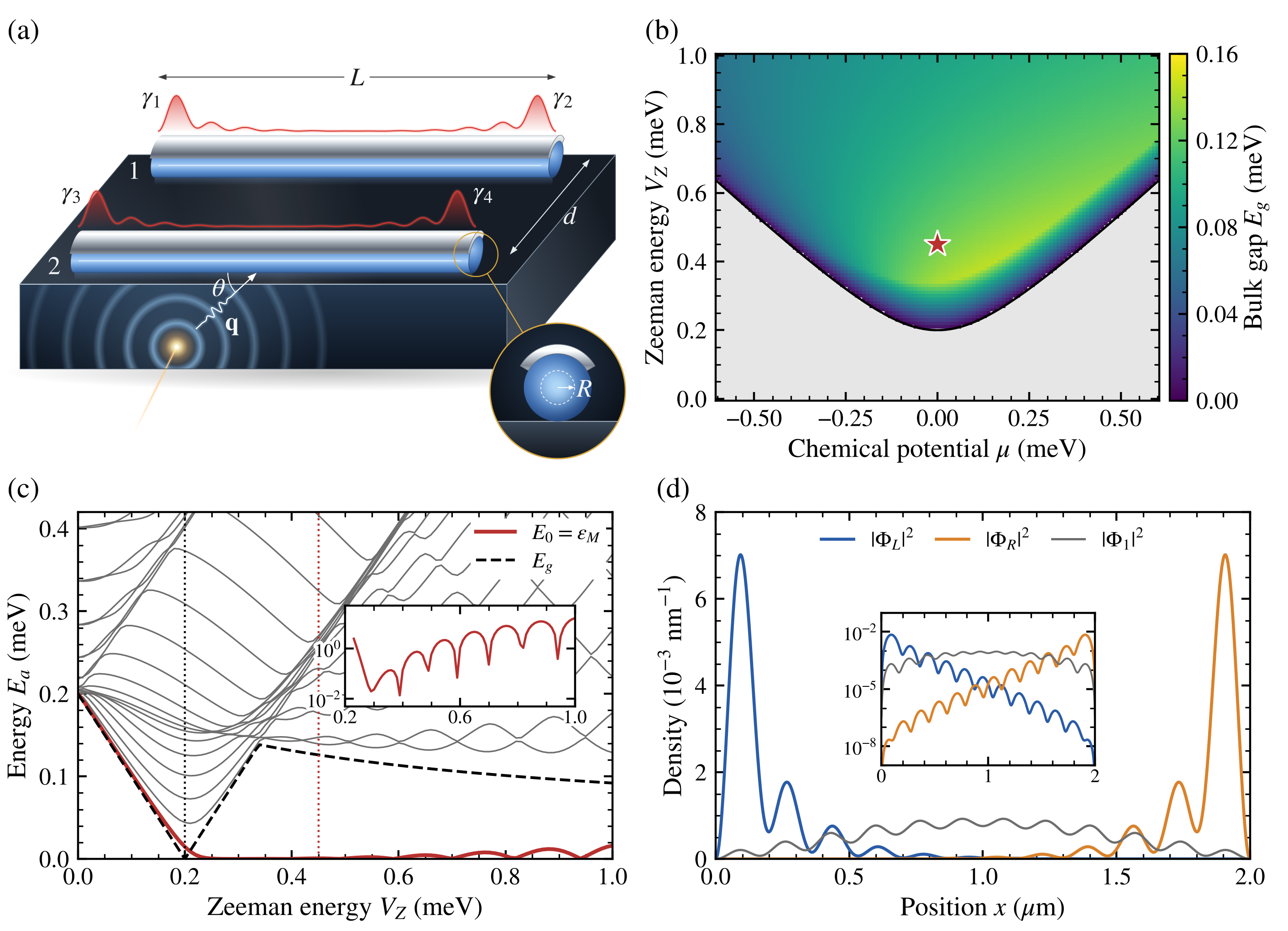}
\caption{Qubit geometry and wire spectrum. (a) Illustration of two parallel semiconductor wires (blue), each coupled to a superconducting shell (gray), host four Majorana end modes $\gamma_1,\dots,\gamma_4$. The wires have length $L$ and separation $d$. A phonon generated by an radiation impact in the substrate reaches a wire at angle $\theta$ to its axis. (b) Bulk gap $E_g$ of the infinite lattice wire in the topological phase, $\mathcal{Q}=-1$. Gray marks the trivial phase, the black line is $V_Z=\sqrt{\mu^2+\Delta^2}$, and the star marks the working point. (c) Positive eigenenergies of the finite $2~\mu\mathrm{m}$ wire versus $V_Z$ at $\mu=0$. Gray lines show the excited modes, red shows the lowest energy $E_0=\eps$, and the dashed line shows the bulk gap. Vertical dotted lines mark $V_Z=\Delta$ and the working point. The inset shows $\eps$ in $\mu$eV on a logarithmic scale. (d) Spatial densities of the Majorana spinors $\Phi_L$ and $\Phi_R$ and the lowest bulk mode $\Phi_1$ at the working point. The inset shows the same densities in nm$^{-1}$ on a logarithmic scale versus $x$ in $\mu$m.}
\label{fig:spectrum}
\end{figure*}

We consider two identical parallel wires separated by $d=300$~nm, with no tunneling between them [Fig.~\ref{fig:spectrum}(a)]. Wire 1 hosts the Majorana end modes $\gamma_1$ and $\gamma_2$, and wire 2 hosts $\gamma_3$ and $\gamma_4$. The two Majorana modes in each wire form one fermionic mode with occupation $n_0=0$ or 1. We use the even-total-parity states
\begin{equation}
|0_L\rangle=|00\rangle,\qquad |1_L\rangle=|11\rangle
\label{eq:logical-basis}
\end{equation}
as the logical basis, with all bulk quasiparticle modes empty. Here the two entries specify the fermionic occupations in wires 1 and 2. The operators $X_L$, $Y_L$, and $Z_L$ are the Pauli operators in this basis, and the off-diagonal terms in the density matrix $\rho_{01}=\langle0_L|\rho|1_L\rangle$ measures the logical coherence. Bulk quasiparticles in either wire cause leakage from the logical subspace. Quasiparticle exchange with an external reservoir can leave a wire’s Majorana occupation flipped even after the bulk empties, resulting in poisoning.

Each wire is described by the Lutchyn--Oreg model~\cite{Lutchyn2010,Oreg2010}, which combines a semiconductor band, Rashba spin--orbit coupling $\alpha$, a Zeeman energy $V_Z$ along the wire, and a fixed induced pairing $\Delta$. We introduce the four-component Nambu field operator
\begin{equation}
\Psi=(\psi_\uparrow,\psi_\downarrow,
\psi_\downarrow^\dagger,-\psi_\uparrow^\dagger)^T,
\label{eq:nambu-spinor}
\end{equation}
where $\psi_\sigma(x)$ annihilates an electron with spin $\sigma$ at position $x$, and $\psi_\sigma^\dagger(x)$ creates one. In this time-reversed-hole basis, the Hamiltonian is
\begin{equation}
H_{\rm LO}
=\left(-\frac{\hbar^2\partial_x^2}{2m^*}-\mu\right)\tau_z
-i\alpha\partial_x\sigma_y\tau_z
+V_Z\sigma_x+\Delta\tau_x,
\label{eq:HLO}
\end{equation}
where $m^*$ is the effective mass, $\mu$ is the chemical potential, and $\sigma_i$ and $\tau_i$ are Pauli matrices acting in spin and particle--hole space, respectively.

We discretize the wire with lattice spacing $a$ and define $t_0=\hbar^2/(2m^*a^2)$. The on-site and nearest-neighbor blocks are
\begin{equation}
\begin{aligned}
H_{ii}&=(2t_0-\mu)\tau_z+V_Z\sigma_x+\Delta\tau_x,\\
H_{i,i+1}&=-t_0\tau_z-\frac{i\alpha}{2a}\sigma_y\tau_z.
\end{aligned}
\label{eq:Hlat}
\end{equation}
There are $L/a-1$ interior sites, with hard-wall boundaries at $x=0$ and $x=L$. In this representation $H$ is real, and particle--hole conjugation is $\mathcal{C}=\tau_y\sigma_yK$, where $K$ denotes complex conjugation.

The lattice topological invariant is
\begin{equation}
\begin{aligned}
\mathcal{Q}={}&\mathrm{sgn}\!\big[
(\mu^2+\Delta^2-V_Z^2)\\
&\times\big((4t_0-\mu)^2+\Delta^2-V_Z^2\big)
\big].
\end{aligned}
\label{eq:topological-invariant}
\end{equation}
The topological phase has $\mathcal{Q}=-1$. At the low fillings considered here ($|\mu|,\Delta,V_Z\ll4t_0$), this condition reduces to $V_Z>\sqrt{\mu^2+\Delta^2}$.

Diagonalizing the finite-wire Hamiltonian gives the positive-energy modes $\Phi_a$ and energies $E_a$. The lowest mode, $\Phi_0$, has energy $E_0\equiv\eps\ge0$. It is formed from the two end-localized, self-conjugate spinors $\Phi_L$ and $\Phi_R$, for which $\mathcal{C}\Phi_{L,R}=\Phi_{L,R}$. Their overlap produces the finite splitting $\eps$. We refer to the modes $a>0$ as bulk modes and retain all positive-energy modes below 1.2~meV.

We use InAs-like effective parameters: $m^*=0.023\,m_e$, $\alpha=20~\mathrm{meV\,nm}$, $\mu=0$, $V_Z=0.45~\mathrm{meV}$, and $\Delta=0.20~\mathrm{meV}$. Each wire is $2~\mu\mathrm{m}$ long and is discretized with spacing $a=2.5$~nm. The phonon bath is at 20~mK, and the pulse has a Gaussian time envelope with a full width at half maximum of 50~ns. These values define the reference working point; the remaining material inputs, pulse spectra, and numerical settings are collected in Appendix~\ref{app:parameters}.

\subsection{Electron-phonon coupling}

A phonon mode \(\lambda\) produces a local potential \(g_\lambda(x)\) in the semiconductor. Because this potential shifts electron and hole energies with opposite signs, it adds the perturbation \(V_\lambda=g_\lambda(x)\tau_z\) to \(H_{\rm LO}\). Expressing this perturbation in the quasiparticle basis obtained by diagonalizing \(H_{\rm LO}\) gives the coupling operator
\begin{equation}
O_\lambda=
\sum_{ab}A^\lambda_{ab}c_a^\dagger c_b
+\sum_{a<b}\left(B^\lambda_{ab}c_a^\dagger c_b^\dagger
+{\rm h.c.}\right),
\label{eq:vertex}
\end{equation}
with matrix elements
\begin{equation}
\begin{aligned}
A^\lambda_{ab}&=\langle\Phi_a|V_\lambda|\Phi_b\rangle,\\
B^\lambda_{ab}&=\langle\Phi_a|V_\lambda\mathcal{C}|\Phi_b\rangle.
\end{aligned}
\label{eq:coupling-matrix-elements}
\end{equation}
The off-diagonal $A$ terms describe phonon-mediated transitions of a quasiparticle from mode $b$ to mode $a$, while the diagonal terms shift the energy of an occupied mode. The $B$ terms create or annihilate a pair. For real standing-wave phonons, $A=A^\dagger$ and $B=-B^T$, so $B_{aa}=0$: a pair cannot occupy the same quasiparticle mode. For the incoherent, isotropic phonon bath considered here, real standing-wave modes and traveling-wave modes give identical total transition rates. We can therefore use either representation.

We model phonons as plane waves in an isotropic elastic continuum. Longitudinal acoustic (LA) phonons couple through the deformation potential, with amplitude $iDq$, where $D$ is deformation-potential constant and $q=|\mathbf{q}|$. In zincblende InAs, both LA and transverse acoustic (TA) phonons also produce a piezoelectric potential. Its amplitude is $P\,\mathbf f(\hat{\mathbf q})\cdot\mathbf e$, where
\begin{equation}
\begin{aligned}
\mathbf f&=(\hat q_y\hat q_z,\hat q_z\hat q_x,\hat q_x\hat q_y),\\
P&=\frac{2ee_{14}}{\varepsilon_0\varepsilon_r}
=671~\mathrm{meV\,nm^{-1}}.
\end{aligned}
\label{eq:piezoelectric-amplitude}
\end{equation}
Here $\mathbf e$ is the polarization vector, the components of $\hat{\mathbf q}$ are expressed in crystal axes, $e$ is the positive elementary charge, and $e_{14}$ is the piezoelectric coefficient.

The transverse electron density is Gaussian with $1/e$ radius $R$. Its form factor is
\begin{equation}
|F_\perp(q_\perp)|^2=e^{-q_\perp^2R^2/2},
\label{eq:transverse-form-factor}
\end{equation}
which suppresses coupling to phonons with transverse wave vector much larger than $1/R$.

To obtain transition rates from the coupling matrix elements
$A_{ab}$ and $B_{ab}$, we first separate the electronic spatial
overlap from the phonon coupling factors. For a phonon whose
wave-vector component along the wire is $k$, the electronic
factor is
\begin{equation}
M_{ab}(k)=\sum_i e^{ikx_i}\rho_{ab}(x_i),
\label{eq:transition-form-factor}
\end{equation}
where
\begin{equation}
\rho_{ab}(x_i)=
\begin{cases}
\Phi_a^\dagger(x_i)\tau_z\Phi_b(x_i),
& \text{scattering},\\
\Phi_a^\dagger(x_i)\tau_z\mathcal{C}\Phi_b(x_i),
& \text{pair creation}.
\end{cases}
\label{eq:transition-density}
\end{equation}
Thus, $M_{ab}$ is the electronic part of the corresponding
$A_{ab}$ or $B_{ab}$ matrix element. It determines how strongly
a phonon with a given longitudinal wave vector couples the
two quasiparticle modes.

Applying Fermi's golden rule and integrating over phonon
propagation directions gives the transition rate per unit
phonon occupation:
\begin{multline}
\gamma^\nu_{ab}(\Omega)=
\frac{q}{8\pi^2\hbar\rho c_\nu^2}
\int_{-1}^{1}\!d\cos\theta\,\bar C_\nu(\theta,q)\\
\times |F_\perp(q\sin\theta)|^2
|M_{ab}(q\cos\theta)|^2,
\label{eq:rate}
\end{multline}
where $\Omega>0$ is the energy exchanged with the phonon,
$\rho$ is the material mass density, and $c_\nu$ is the sound
velocity of branch $\nu$. Energy conservation fixes
$q=\Omega/(\hbar c_\nu)$, and $\theta$ is the angle between
the phonon wave vector and the wire axis. The factors
$F_\perp(q\sin\theta)$ and $M_{ab}(q\cos\theta)$ account for
the transverse and longitudinal spatial overlaps,
respectively.

The quantity $\bar C_\nu$ is obtained by adding the coupling
amplitudes for each phonon mode, taking the squared magnitude,
and then integrating over the azimuthal angle $\phi$:
\begin{multline}
\bar C_\nu(\theta,q)
=\int_0^{2\pi}\!d\phi\sum_{s=1}^{g_\nu}\\
\times\left|
iDq\,\delta_{\nu,\mathrm{LA}}
+P\,\mathbf f(\hat{\mathbf q})\cdot\mathbf e_{\nu s}
\right|^2.
\label{eq:angular-coupling-factor}
\end{multline}
Here $s$ labels the polarization, with $g_{\rm LA}=1$ and
$g_{\rm TA}=2$, and $\delta_{\nu,\mathrm{LA}}$ restricts
deformation-potential coupling to LA phonons. The phonon
direction and polarization are expressed in crystal axes
when evaluating $\mathbf f\cdot\mathbf e_{\nu s}$.

For an LA phonon, both mechanisms contribute to the same
mode. With real polarization vectors, the deformation-potential
amplitude $iDq$ is purely imaginary, whereas the piezoelectric
amplitude $P\,\mathbf f\cdot\mathbf e_{\rm LA}$ is real.
Their interference term,
$2\operatorname{Re}[(iDq)
(P\,\mathbf f\cdot\mathbf e_{\rm LA})^*]$, therefore vanishes.
The squared total amplitude consequently separates into
deformation-potential and piezoelectric contributions:
\begin{equation}
\bar C_{\rm LA}^{\rm DP}=2\pi D^2q^2,
\label{eq:deformation-angular-factor}
\end{equation}
and
\begin{equation}
\bar C_\nu^{\rm PE}
=P^2\int_0^{2\pi}\!d\phi
\sum_{s=1}^{g_\nu}
\left|\mathbf f(\hat{\mathbf q})\cdot\mathbf e_{\nu s}\right|^2.
\label{eq:piezoelectric-angular-factor}
\end{equation}
Thus,
$\bar C_{\rm LA}=\bar C_{\rm LA}^{\rm DP}
+\bar C_{\rm LA}^{\rm PE}$, while
$\bar C_{\rm TA}=\bar C_{\rm TA}^{\rm PE}$.

The quantity $\gamma^\nu_{ab}$ contains the microscopic
coupling strength but not the phonon occupation. Multiplying
it by $N_\nu(\Omega,t)$ gives the absorption rate, while
multiplying it by $N_\nu(\Omega,t)+1$ gives the emission rate.
After summing over branches, these rates enter the Pauli
master equation for the quasiparticle occupation
configurations. The same $\gamma^\nu_{ab}$ values provide
the transition-line weights in the excitation kernel
defined below.

\subsection{Excitation kernel and pulses}

Starting with an empty bulk, a phonon can excite the wire through three channels. If the Majorana occupation is $p=n_0=0$, it can create a pair consisting of one excitation in the Majorana fermionic mode and one in bulk mode $a$. This process has weight $|B_{0a}|^2$ and requires energy $E_a+\eps$. If $p=1$, it can instead promote the existing quasiparticle into bulk mode $a$, with weight $|A_{a0}|^2$ and energy $E_a-\eps$. For either value of $p$, it can also create two bulk quasiparticles, with weight $|B_{ab}|^2$ and energy $E_a+E_b$, where $0<a<b$.

We collect these discrete transitions into the excitation kernel
\begin{equation}
\Kp{p}_\nu(\Omega)=
\sum_\ell\gamma^\nu_\ell\delta(\Omega-\Omega_\ell),
\label{eq:kernel}
\end{equation}
where $\ell$ labels an individual allowed excitation for the specified initial Majorana occupation $p$. This kernel gives the excitation rate per unit phonon occupation and unit phonon energy. For an isotropic occupation $N_\nu(\Omega,t)$, the total rate is
\begin{equation}
\Gamma^{(p)}_{\rm exc}(t)=
\sum_\nu\int d\Omega\,
\Kp{p}_\nu(\Omega)N_\nu(\Omega,t).
\label{eq:excitation-rate}
\end{equation}
For plotting, we broaden each line with a Gaussian of the width specified in Appendix~\ref{app:parameters}.

A pulse adds the excess occupation
\begin{equation}
\delta N(\Omega,t)=\mathcal A\,s(t)\,\pi(\Omega)
\label{eq:pulse-occupation}
\end{equation}
to the 20-mK bath. We use the same excess occupation for all phonon branches. The envelope $s(t)$ sets the time dependence, while $\pi(\Omega)$ sets the spectrum. We consider a narrowband Gaussian centered at $\Omega_c$, a broadband step, and a thermal excess $n_B(\Omega,T^*)-n_B(\Omega,T_{\rm bath})$ as comparison.

The amplitude $\mathcal A$ fixes the peak excess phonon energy density,
\begin{equation}
\begin{aligned}
u&=\sum_\nu\int d\Omega\,
\frac{F_\nu}{V}\,\Omega\,\delta N_\nu(\Omega),\\
\frac{F_\nu}{V}&=
\frac{g_\nu\Omega^2}{2\pi^2\hbar^3c_\nu^3},
\end{aligned}\label{eq:dose}
\end{equation}
where $F_\nu/V$ is the phonon density of states per unit volume, with $g_{\rm LA}=1$ and $g_{\rm TA}=2$. Here $\delta N_\nu(\Omega)$ denotes the occupation at the pulse peak. We compare pulses with different phonon spectra but the same
peak excess energy density $u$ and temporal envelope $s(t)$
to determine how the spectrum affects excitation probabilities
and logical coherence loss. Their weak-pulse excitation efficiencies are
\begin{equation}
\eta=\lim_{u\to0}\frac{\Pexc}{u},
\label{eq:excitation-efficiency}
\end{equation}
Here $\Pexc\equiv P_{\rm exc}^{(0)}$ is the probability of at least one excitation during the pulse, starting with an empty Majorana level and no bulk quasiparticles [Eq.~\eqref{eq:excitation-probability}]. Thus $\eta$ measures excitation probability per unit excess energy density in the weak-pulse limit.

\subsection{Parity-resolved kinetics}

A weak pulse creates only a few quasiparticles, whose occupations can remain correlated. We therefore evolve the probability $P_n$ of each many-body configuration
\begin{equation}
|n\rangle=\prod_a(c_a^\dagger)^{n_a}|{\rm vac}\rangle,
\qquad n_a=0,1,
\label{eq:occupation-configuration}
\end{equation}
using the configuration-resolved Pauli master equation. We first define the single-process rates. For $\Omega_{ab}=|E_a-E_b|>0$, scattering has rate
\begin{equation}
W_{a\leftarrow b}(t)=
\sum_\nu\gamma_{ab}^{\nu,A}(\Omega_{ab})
\begin{cases}
N_\nu(\Omega_{ab},t),&E_a>E_b,\\
N_\nu(\Omega_{ab},t)+1,&E_a<E_b.
\end{cases}
\label{eq:scattering-process-rate}
\end{equation}
Pair creation and recombination have rates
\begin{equation}
\begin{aligned}
W^+_{ab}(t)&=\sum_\nu
\gamma_{ab}^{\nu,B}(E_a+E_b)N_\nu(E_a+E_b,t),\\
W^-_{ab}(t)&=\sum_\nu
\gamma_{ab}^{\nu,B}(E_a+E_b)[N_\nu(E_a+E_b,t)+1].
\end{aligned}
\label{eq:pair-process-rates}
\end{equation}
Here the superscripts $A$ and $B$ specify the scattering or pair-creation form factor in Eq.~\eqref{eq:rate}; all $W$ have units of inverse time and exclude electronic occupation factors.

Let $R_{m\leftarrow n}$ denote the rate from configuration $n$ to $m$, and let $\mathbf e_a$ change occupation $n_a$ by one. The nonzero phonon-induced rates are
\begin{equation}
\begin{aligned}
R_{n+\mathbf e_a-\mathbf e_b\leftarrow n}
&=W_{a\leftarrow b}n_b(1-n_a),&&a\ne b,\\
R_{n+\mathbf e_a+\mathbf e_b\leftarrow n}
&=W^+_{ab}(1-n_a)(1-n_b),&&a<b,\\
R_{n-\mathbf e_a-\mathbf e_b\leftarrow n}
&=W^-_{ab}n_an_b,&&a<b.
\end{aligned}
\label{eq:configuration-transition-rates}
\end{equation}
The occupation factors permit only valid fermionic configurations. Their probabilities obey
\begin{equation}
\dot P_n(t)=\sum_{m\ne n}
\left[R_{n\leftarrow m}(t)P_m(t)
-R_{m\leftarrow n}(t)P_n(t)\right].
\label{eq:pauli-configuration-master}
\end{equation}
Equivalently, $\dot{\mathbf P}=\mathcal L(t)\mathbf P$, with
\begin{equation}
\mathcal L_{nm}=
\begin{cases}
R_{n\leftarrow m},&n\ne m,\\
-\displaystyle\sum_{k\ne n}R_{k\leftarrow n},&n=m.
\end{cases}
\label{eq:pauli-generator-elements}
\end{equation}
Every phonon-induced jump changes the quasiparticle number by zero or two. Thus $R_{m\leftarrow n}=0$ between configurations of opposite parity, $\pi_n=(-1)^{\sum_a n_a}$, and Eq.~\eqref{eq:pauli-configuration-master} evolves the even and odd sectors separately. An initially empty bulk with Majorana occupation $p$ starts in the sector $\pi=(-1)^p$.

The logical subspace is spanned by
$|0_L\rangle=|00\rangle$ and $|1_L\rangle=|11\rangle$,
where the two entries denote the Majorana-level occupations
of the two wires and all bulk quasiparticle modes are empty.
A bulk excitation in either wire takes the qubit outside
this subspace.

For each wire, we retain configurations with at most
$k_{\max}$ bulk quasiparticles. We monitor probability
flowing beyond this cutoff separately as a measure of
numerical truncation error. Within the retained space,
the joint configuration probabilities preserve correlations
between all mode occupations.

We monitor the bulk leakage probability, which measures
whether any bulk mode is occupied, and the probability
that the Majorana occupation differs from its initial value:
\begin{equation}
\begin{aligned}
\Pleak&=1-\sum_n\prod_{a>0}(1-n_a)P_n,\\
\Pflip&=\sum_{n:\,n_0\neq p}P_n,
\end{aligned}
\label{eq:obs}
\end{equation}
where $p$ is the initial Majorana occupation. Here the Majorana occupation change described by $\Pflip$ can occur transiently without changing the total wire parity. 

Starting with an empty bulk and Majorana occupation $p$,
we calculate the probability that the
pulse produces at least one bulk excitation, regardless
of whether the bulk subsequently returns to its empty state:
\begin{equation}
P^{(p)}_{\rm exc}
=1-\exp\!\left[-\int dt\,\Gamma^{(p)}_{\rm exc}(t)\right].
\label{eq:excitation-probability}
\end{equation}
We write $\Pexc$ for the case $p=0$. To characterize recovery, we calculate the mean time required for the bulk to empty,
\begin{equation}
\tret=-\mathbf1^T\mathcal L_b^{-1}\mathbf P_*.
\label{eq:return-time}
\end{equation}
Here $\mathcal L_b$ contains only the equilibrium-bath transitions and is restricted to configurations with a nonempty bulk. We evaluate the post-pulse distribution $\mathbf P_*$
at $t_*=t_c+6\sigma_t$, where $t_c$ is the pulse-center
time and $\sigma_t$ is the temporal standard deviation
of the Gaussian envelope. By this time, the pulse has
effectively ended. We include only configurations that
still contain bulk quasiparticles and normalize their
probabilities to sum to one. An excitation and its subsequent return to the empty bulk together form a full excursion process out of the code space.

For comparison, we solve the factorized Boltzmann equation using the same transition-rate tables. This approximation replaces joint occupation averages by products, $\langle n_an_b\rangle=f_af_b$, where $f_a=\langle n_a\rangle$. It gives
\begin{multline}
\dot f_a=
\sum_{b\neq a}\left[
W_{a\leftarrow b}f_b(1-f_a)
-W_{b\leftarrow a}f_a(1-f_b)\right]\\
+\sum_{b\neq a}\left[
W^+_{ab}(1-f_a)(1-f_b)-W^-_{ab}f_af_b\right].
\label{eq:boltz}
\end{multline}
Because the microscopic rates are identical in the two calculations, their differences isolate the effect of neglecting occupation correlations.

We also consider escape of bulk quasiparticles to an external reservoir at a phenomenological rate $\Gesc$. In $\mathcal L$, escape is a jump that empties an occupied bulk mode. In Eq.~\eqref{eq:boltz}, it adds $-\Gesc f_a$ for $a>0$. Each escape event changes the parity of the wire.
Its contribution to Eq.~\eqref{eq:pauli-configuration-master} is
\begin{equation}
R^{\rm esc}_{n-\mathbf e_a\leftarrow n}=\Gesc n_a,
\qquad a>0,
\label{eq:escape-configuration-rate}
\end{equation}
which couples the even- and odd-parity sectors.

\subsection{Logical coherence}

Excitation can carry coherence outside the code space,
and relaxation can return it when the bulk empties.
To distinguish this temporary transfer from irreversible
coherence loss, we supplement the population dynamics
with selected off-diagonal elements of the many-body
density matrix, evolved under a Born--Markov master
equation with the same coupling vertices~\cite{Breuer2002}.

We label configurations by $|n_0,S\rangle$, where $n_0$
is the Majorana occupation and $S$ lists the occupied
bulk modes; the partner wire's Majorana occupation is
implicit in these labels. We define
\begin{equation}
\chi_S(t)=
\frac{\langle0,S|\rho(t)|1,S\rangle}
{\langle0,\varnothing|\rho(0)|1,\varnothing\rangle}.
\label{eq:cS}
\end{equation}
For each fixed configuration $S$, $\chi_S(t)$ tracks
the coherence between the two Majorana occupations relative to the initial empty-bulk coherence.
The quantity $\chi_\varnothing$ measures coherence in
the empty-bulk sector. Transitions involving bulk modes alone leave $n_0$
unchanged and act identically on the two branches.
They transfer coherence from $\chi_S$ to $\chi_{S'}$
at the same rate as the corresponding population
transition. Subsequent processes involving the Majorana
level can, however, allow phonons to distinguish the
branches and destroy this coherence. Tracking the
$\chi_S$ therefore determines how much coherence returns
to the empty-bulk sector rather than being lost during
the excursion.

We evolve one wire while treating the other as a
spectator. The partner wire's Majorana occupation remains
an implicit label of the two logical branches; these
matrix elements therefore retain the two-wire logical
coherence. Escape acts only on the active wire, but can change
the total parity of the two-wire system. We therefore
retain the partner wire's Majorana occupation as an
independent state label, rather than infer it from
the fixed-parity constraint.

Transitions that change $n_0$ can distinguish the logical
branches: pair creation exchanges a phonon of energy
$E_a+\eps$, whereas promotion exchanges one of energy
$E_a-\eps$. These processes interchange the Majorana
occupations, producing damping at the mean of the
branch rates and coherence transfer between configurations
$S$ and $S\pm a$, where $S\pm a$ denotes adding or removing
bulk mode $a$. The matrix-element equations below fix
the transfer signs and conjugations explicitly.

To compare their phonon mode profiles, we calculate
the promotion rate coefficient $\Gamma_a^A$, the
pair-creation rate coefficient $\Gamma_a^B$, and the
coherence-transfer coefficient $\Gamma_a^\times$
at the common energy $\Omega=E_a$, before including
phonon occupation factors.
The direct coefficients are
\begin{equation}
\Gamma_a^A=\sum_\nu\gamma_{a0}^{\nu,A}(E_a),
\qquad
\Gamma_a^B=\sum_\nu\gamma_{a0}^{\nu,B}(E_a),
\label{eq:direct-coherence-coefficients}
\end{equation}
where Eq.~\eqref{eq:rate} uses the electronic form
factors $M^A_{a0}$ for promotion and $M^B_{a0}$ for
pair creation, respectively. The cross coefficient
contains their amplitude product and governs coherence
transfer:
\begin{multline}
\Gamma^\times_a=
\sum_\nu\frac{q_\nu}{8\pi^2\hbar\rho c_\nu^2}
\int_{-1}^{1}\!d\cos\theta\,
\bar C_\nu(\theta,q_\nu)
|F_\perp(q_\nu\sin\theta)|^2\\
\times M^B_{a0}(q_\nu\cos\theta)
M^A_{a0}(q_\nu\cos\theta)^*,
\label{eq:cross}
\end{multline}
with $q_\nu=E_a/(\hbar c_\nu)$.

The normalized overlap is
\begin{equation}
\kappa_a=
\frac{|\Gamma^\times_a|}
{\sqrt{\Gamma_a^A\Gamma_a^B}}\le1.
\label{eq:phonon-overlap}
\end{equation}
At this common energy, $\kappa_a=0$ means that the
phonon states are orthogonal and distinguish the
branches, while $\kappa_a=1$ means that their normalized
mode profiles are identical up to a phase.
For absorption or emission, the corresponding kinetic
coefficients are obtained by multiplying each branch
contribution to all three coefficients by
$N_\nu(E_a,t)$ or $N_\nu(E_a,t)+1$, respectively.
Note that vanishing $\kappa_a$ does not imply complete qubit dephasing, since the pulse-averaged loss also depends on the probability of taking the corresponding excitation paths (Appendix~\ref{app:phonon-overlap}).

The explicit coherence equations use the same fermionic transition operators as the population kinetics. For $a>0$, define the pair-creation and promotion operators
\begin{equation}
F_{aB}^{+}=c_a^\dagger c_0^\dagger,\qquad
F_{aA}^{+}=c_a^\dagger c_0,\qquad
F_{aX}^{-}=(F_{aX}^{+})^\dagger.
\label{eq:paired-jump-operators}
\end{equation}
The order $c_a^\dagger c_0^\dagger$ matches the vertex $B_{a0}$ in Eq.~\eqref{eq:cross}. The absorption and emission blocks, ordered as $(B,A)$, are
\begin{equation}
\begin{aligned}
\mathsf C_a^+&=
\begin{pmatrix}
d_{aB}^+&r_a^+\\
(r_a^+)^*&d_{aA}^+
\end{pmatrix},\\
\mathsf C_a^-&=
\begin{pmatrix}
d_{aB}^-&(r_a^-)^*\\
r_a^-&d_{aA}^-
\end{pmatrix},
\end{aligned}
\label{eq:paired-rate-matrices}
\end{equation}
where
\begin{equation}
\begin{aligned}
(d_{aB}^+,d_{aA}^+)&=(W^+_{0a},W_{a\leftarrow0}),\\
(d_{aB}^-,d_{aA}^-)&=(W^-_{0a},W_{0\leftarrow a}).
\end{aligned}
\label{eq:paired-direct-rates}
\end{equation}
Writing $\gamma_a^{\nu,\times}$ for the contribution of branch $\nu$ to Eq.~\eqref{eq:cross}, the common-energy cross coefficients are
\begin{equation}
\begin{aligned}
r_a^+(t)&=\sum_\nu N_\nu(E_a,t)\gamma_a^{\nu,\times},\\
r_a^-(t)&=\sum_\nu[N_\nu(E_a,t)+1]\gamma_a^{\nu,\times}.
\end{aligned}
\label{eq:paired-cross-rates}
\end{equation}
The conjugation in the emission block follows from reversing the transition operators and their microscopic amplitudes.

Let $\alpha$ label these two-channel blocks and the remaining single-channel jumps. The latter are bulk scattering $c_a^\dagger c_b$ with rate $W_{a\leftarrow b}$, bulk-pair creation $c_a^\dagger c_b^\dagger$ and recombination with rates $W^+_{ab}$ and $W^-_{ab}$, and escape $c_a$ with rate $\Gesc$; the bulk-only indices satisfy $a,b>0$, with $a\ne b$ for scattering and $a<b$ for pair processes. A single-channel block has its rate as its sole matrix entry. In a frame removing the spectator wire's known free precession, the master equation is
\begin{multline}
\dot\rho=-\frac{i}{\hbar}[H_{\rm qp},\rho]\\
+\sum_{\alpha,ij}(\mathsf C_\alpha)_{ij}
\left(F_{\alpha i}\rho F_{\alpha j}^\dagger
-\frac12\{F_{\alpha j}^\dagger F_{\alpha i},\rho\}\right),
\label{eq:coherence-operator-master}
\end{multline}
where $H_{\rm qp}=\sum_aE_ac_a^\dagger c_a$ acts on the active wire, while $\rho$ retains both wires' state labels. The Hamiltonian Lamb shift is omitted.

For an explicit equation for the retained matrix elements, write $\rho_{nm}=\langle n|\rho|m\rangle$, with the spectator occupation included in $n,m$, and define
\begin{multline}
K_{nm;kl}=\sum_{\alpha,ij}(\mathsf C_\alpha)_{ij}\\
\times\langle n|F_{\alpha i}|k\rangle
\langle m|F_{\alpha j}|l\rangle^*.
\label{eq:coherence-transfer-elements}
\end{multline}
For these transition blocks the loss operator is diagonal in the occupation basis. With $E_n=\sum_aE_an_a$ and total outgoing rate $\Lambda_n=\sum_{k\ne n}K_{kk;nn}$, the component equation is
\begin{multline}
\dot\rho_{nm}=
\left[-\frac{i(E_n-E_m)}{\hbar}
-\frac{\Lambda_n+\Lambda_m}{2}\right]\rho_{nm}\\
+\sum_{kl}K_{nm;kl}\rho_{kl}.
\label{eq:coherence-component-master}
\end{multline}
For populations, $K_{nn;kk}=R_{n\leftarrow k}$, recovering Eq.~\eqref{eq:pauli-configuration-master}. The off-diagonal equation transfers coherence between configurations and damps it at the mean outgoing rate.
Let $n_S$ and $m_S$ label the two many-body configurations
whose coherence is tracked by $\chi_S$. We write each
configuration as $(n_0,S;n_{\rm sp})$, where $n_0$ is the
active wire's Majorana occupation, $S$ lists its occupied
bulk modes, and $n_{\rm sp}$ is the spectator wire's
Majorana occupation. The spectator bulk remains empty.

Without escape, the total parity remains even, so
\begin{equation}
\begin{aligned}
n_S&=(0,S;r_S),\qquad m_S=(1,S;1-r_S),\\
r_S&=|S|\bmod2.
\end{aligned}
\label{eq:coherence-paired-state-labels}
\end{equation}
Here $|S|$ counts the bulk quasiparticles. Thus $r_S=0$
for an even number and $r_S=1$ for an odd number; the
spectator occupations shown above ensure even total
parity in both configurations. For $C_0=\rho_{01}(0)\ne0$, the closed same-bulk-configuration equations are
\begin{multline}
\dot\chi_S=
\left[\frac{i\eps}{\hbar}
-\frac{\Lambda_{n_S}+\Lambda_{m_S}}2\right]\chi_S\\
+\sum_T K_{n_Sm_S;n_Tm_T}\chi_T\\
+\frac{C_0^*}{C_0}\sum_T
K_{n_Sm_S;m_Tn_T}\chi_T^*.
\label{eq:logical-coherence-master}
\end{multline}
The phase factor is unity for real initial coherence; retaining it permits an arbitrary initial logical phase. Additional coherences between different bulk configurations are omitted in this closure. With escape, the spectator occupation is independent and Eq.~\eqref{eq:coherence-component-master} is used with the full state labels. The fermionic matrix elements in Eq.~\eqref{eq:coherence-transfer-elements} fix all transfer signs, including those associated with escape.

Complete positivity is ensured when each $\mathsf C_\alpha$ is positive semidefinite. This is guaranteed by a consistent common-energy construction, but is not implied by the bare overlap bound when the direct rates use $E_a\pm\eps$ and the cross coefficient uses $E_a$. Appendix~\ref{app:coefficient-matrix} gives the explicit condition, its proof, and the distinction between these rate conventions.

The Majorana splitting provides a separate mechanism for suppressing coherence transfer. Free evolution rotates $\chi_S$ and $\chi_S^*$ at opposite frequencies, $\eps/\hbar$ and $-\eps/\hbar$. When their frequency difference $2\eps/\hbar$ is much larger than the transition rates, the transfer terms oscillate rapidly and average out. We therefore neglect these coherence-transfer terms in this regime.

After the pulse, the generator is constant. We obtain the final coherence, $\chi=\chi_\varnothing(\infty)$, from one linear solve in a frame that removes the free precession. Without escape, and after the bulk has emptied, the pulse acts on the logical state as $\rho_{01}\to\chi\rho_{01}$, a channel diagonal in $Z_L$. For independent events in the two wires, their coherence factors multiply.

The splitting scans in Sec.~\ref{sec:coherence} vary $\eps$ while keeping the wave functions fixed. For transitions involving the Majorana mode, the vertices are evaluated at $\Omega=E_a$, and the final coherence is calculated to first order in the phonon energy density $u$.

\subsection{Parent superconductor}

The fixed-pairing model treats the quasiparticle states as entirely contained in the semiconductor. In a hybrid device, each state has weight in both the semiconductor and the superconducting shell. Since the phonon potential considered here acts in the semiconductor, this distribution changes the excitation matrix elements.

To quantify the correction, we compare the fixed-pairing description with a bare wire $h_w$, with $\Delta=0$, tunnel coupled to a parent superconductor. The parent has gap $\Delta_S$ and an energy-independent normal-state density of states. Integrating out its degrees of freedom gives
\begin{equation}
\Sigma_S(E)=
-\gamma_t\frac{E\tau_0-\Delta_S\tau_x}
{\sqrt{\Delta_S^2-E^2}},
\label{eq:sigma}
\end{equation}
where $\gamma_t$ sets the tunneling strength.

The bound-state energies below the parent gap satisfy
\begin{equation}
[E_a-h_w-\Sigma_S(E_a)]\varphi_a=0.
\label{eq:hybrid-poles}
\end{equation}
We choose $\varphi_a^\dagger\varphi_a=1$. Normalization of the full hybrid state then gives the semiconductor weight $Z_a$ through~\cite{Stanescu2017}
\begin{equation}
Z_a^{-1}=
\varphi_a^\dagger[1-\partial_E\Sigma_S(E_a)]\varphi_a.
\label{eq:semiconductor-weight}
\end{equation}
The semiconductor component of the normalized hybrid state is $w_a=\sqrt{Z_a}\,\varphi_a$. A perturbation $\delta h_w$ acting in the semiconductor therefore has the matrix element
\begin{equation}
A_{ab}=\sqrt{Z_aZ_b}\,
\varphi_a^\dagger\delta h_w\varphi_b.
\label{eq:hybrid-vertex}
\end{equation}
Each transition is consequently weighted by the semiconductor components of both states. We solve for the poles below $0.97\Delta_S$, where the semiconductor weight is already small. Higher-energy bound states approach $\Delta_S$ and have semiconductor weights approaching zero.

For comparison, the usual low-energy approximation expands $\Sigma_S(E)$ around $E=0$:
\begin{equation}
\begin{aligned}
H_{\rm eff}&=Z(h_w+\gamma_t\tau_x),\\
Z&=\left(1+\frac{\gamma_t}{\Delta_S}\right)^{-1}.
\end{aligned}
\label{eq:effective-hamiltonian}
\end{equation}
It evaluates the bare perturbation between unit-normalized eigenvectors of $H_{\rm eff}$. 

\subsection{Kitaev-chain control}\label{sec:kitaev}

To identify which results require the semiconductor band structure, we repeat the excitation-kernel, population, coherence, and parameter-scan calculations for spinless Kitaev chains. The chains have the same length and lattice spacing as the Lutchyn--Oreg wire. They also couple to the same phonons through the same density vertex, $g_\lambda(x)\tau_z$.

In the basis $(c_j,c_j^\dagger)$, the Hamiltonian is
\begin{equation}
\begin{aligned}
H_{jj}&=(2t-\mu_K)\tau_z,\\
H_{j,j+1}&=-t\tau_z+i\Delta_p\tau_y,
\end{aligned}\label{eq:HK}
\end{equation}
with particle--hole conjugation $\tau_xK$. The zero-energy solutions have an envelope $e^{-\kappa x}$ and oscillate at wave vector $k_0$, where
\begin{equation}
\begin{aligned}
\tanh(\kappa a)&=\frac{\Delta_p}{t},\\
\cos(k_0a)&=
\frac{2t-\mu_K}{2\sqrt{t^2-\Delta_p^2}}.
\end{aligned}
\label{eq:kitaev-decay-oscillation}
\end{equation}
Their bulk gap is $E_g=2\Delta_p\sin(k_0a)$. Specifying $E_g$, $\kappa$, and $k_0$ therefore fixes the three chain parameters $t$, $\mu_K$, and $\Delta_p$ in closed form.

We construct two controls with different electronic momentum scales, labeled low- and high-density chains because $k_0$ tracks the Fermi momentum and hence the electron density in the regime considered. The low-density chain matches the wire's gap $E_g$, inverse Majorana decay length $\kappa$, and small oscillation wavevector $k_0$. The high-density chain retains the same $E_g$ and $\kappa$ but uses a larger $k_0$, corresponding to a shorter electronic wavelength and higher band filling. We extract $\kappa$ and $k_0$ from its most slowly decaying oscillatory zero-energy solution,
\begin{equation}
e^{-x/\xi}\cos(k_0x+\phi),\qquad \xi=1/\kappa.
\label{eq:majorana-tail}
\end{equation}
At the working point, this is also the slowest-decaying solution overall and determines the Majorana tail discussed in Sec.~\ref{sec:spectrum}. In the scans of Sec.~\ref{sec:lom}, we repeat the matching at each $V_Z$ and $\mu$, using the gap and the most slowly decaying oscillatory solution at that point.

The high-density chain retains the same $E_g$ and $\kappa$ but sets $k_0a=1.25$. Its oscillation wavelength is approximately five lattice spacings, and $2k_0$ exceeds the wave vectors of all phonons considered here. Comparing the two controls tests how changing the electronic momentum scale affects phonon excitation while keeping the gap and Majorana decay length fixed.
\section{Results}\label{sec:results}

Supplemental Tables~\ref{tab:scales}--\ref{tab:lom} collect the numerical results and checks.

\subsection{Spectrum}\label{sec:spectrum}

The working point lies well inside the topological phase [Fig.~\ref{fig:spectrum}(b)]. The finite wire has a bulk gap of 0.13~meV and a Majorana splitting of $\eps=0.75~\mu$eV [Fig.~\ref{fig:spectrum}(c)]. The end modes decay over $\xi=355$~nm, exhibiting a tail, while the lowest bulk mode extends across the wire [Fig.~\ref{fig:spectrum}(d)]. The Fermi wavelength is about 330~nm, reflecting the low electron density. Spectrum and lattice-convergence checks are listed in Table~\ref{tab:scales}.

\begin{figure*}[t]
\includegraphics[width=\textwidth]{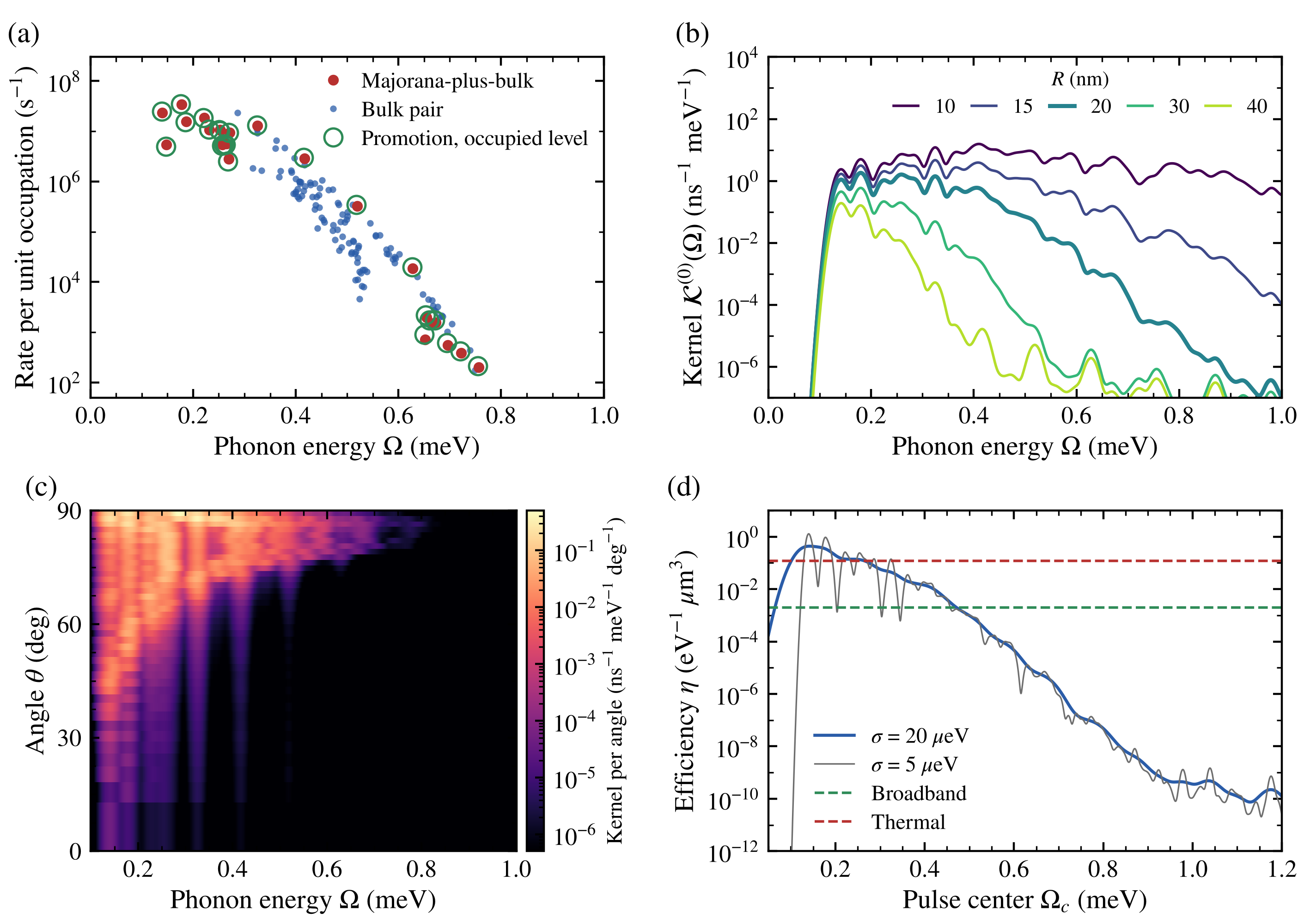}
\caption{Excitation kernel and pulse efficiency at the working point. (a) Cumulative transition rates: total, Majorana-plus-bulk and bulk-pair contributions for an empty Majorana level, and total for an occupied level, normalized to the empty-level total. (b) Kernels for transverse radii $R=10$--40~nm. (c) Angular kernel, with $\theta$ measured from the wire axis and folded onto $0^\circ$--$90^\circ$. (d) Weak-pulse efficiency $\eta$ for narrowband spectra of bandwidth $\sigma$. Dashed lines show $-\ln(1-\Pexc)/u$ for broadband and thermal pulses (Table~\ref{tab:pulses}).}
\label{fig:kernel}
\end{figure*}

\subsection{Phonon excitation}\label{sec:kernelres}
\label{sec:maps}

The lowest allowed transitions create a Majorana-plus-bulk pair; creation of two bulk quasiparticles begins at higher energy. Two thirds of the summed transition rates lie below 0.3~meV [Fig.~\ref{fig:kernel}(a)]. For an occupied Majorana level, promotion into the bulk has similar rates, so the two logical states have similar excitation probabilities (Table~\ref{tab:scales}).

Momentum matching favors phonons traveling nearly perpendicular to the wire [Fig.~\ref{fig:kernel}(c)]. The electronic transition densities carry most of their Fourier weight below about $2k_F$, requiring $q|\cos\theta|\lesssim2k_F$. Most coupling therefore comes from $\theta>70^\circ$, where $q_\perp\approx q$. Transverse confinement suppresses these phonons once $q_\perp R\gtrsim1$, explaining the strong radius dependence in Fig.~\ref{fig:kernel}(b). The summed rates scale approximately as $R^{-4.3}$ over the range studied. Branch-resolved rates, piezoelectric fractions, and field and chemical-potential scans are summarized in Table~\ref{tab:scales}; the different behavior near the two transitions is discussed in Sec.~\ref{sec:lom}.

At equal excess energy density and pulse envelope, narrowband pulses excite the wire most efficiently just above threshold [Fig.~\ref{fig:kernel}(d)]. The efficiency is about 30 times smaller at 0.40~meV and nine orders of magnitude smaller at 1~meV. Both the electronic coupling and the phonon occupation per mode decrease with energy. For a narrow spectrum of fixed width, the latter scales as $u/\Omega_c^3$. The broadband pulse places most of its energy above 0.5~meV and is consequently much less effective than the thermal pulse of equal energy density (Tables~\ref{tab:scales} and \ref{tab:pulses}). Thus deposited energy alone does not determine excitation.

\begin{figure*}[t]
\includegraphics[width=\textwidth]{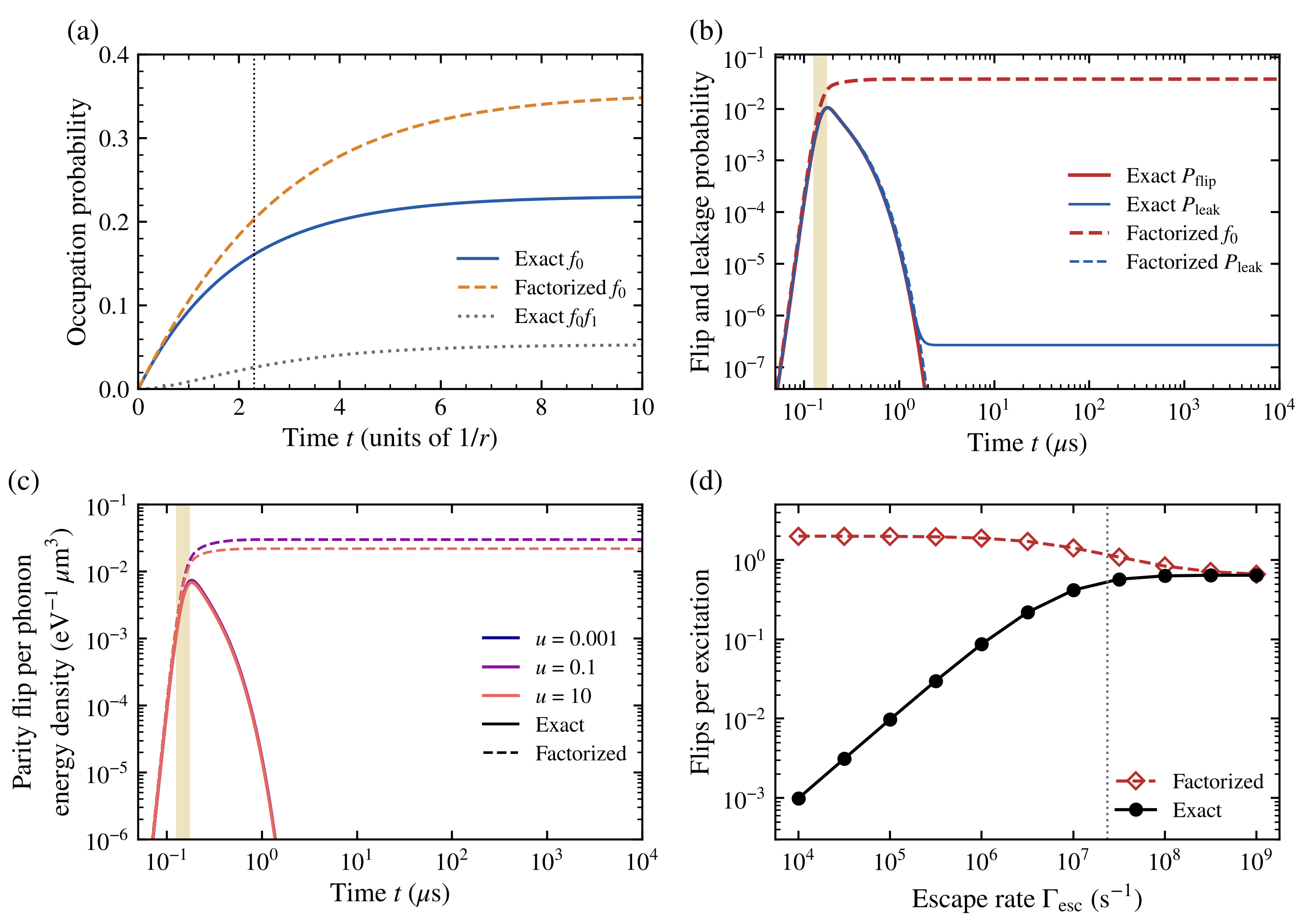}
\caption{Leakage, parity recovery, and poisoning. (a) Two-mode pair dynamics with $r_+=0.12r$ and $r_-=0.40r$: exact occupation $f_0=f_1=P_{11}=\langle n_0n_1\rangle$, factorized occupation, and product of the exact mean occupations. (b) A 0.20-meV pulse at $u=0.1$~\evum{}: exact $\Pflip$ and $\Pleak$ (solid), factorized $f_0$ and leakage (dashed). Shading marks the pulse FWHM; the late leakage plateau is discarded probability beyond $k_{\max}=2$. (c) Exact $\Pflip/u$ and factorized $f_0/u$ for a broadband 0.1--0.6~meV pulse at three energy densities (17 modes, $k_{\max}=4$). (d) Residual flips per excitation at 10~ms versus escape rate for the broadband 0.1--1.0~meV pulse at the reference energy density. The dotted line marks the lowest pair's recombination rate, $2.35\times10^7$~s$^{-1}$.}
\label{fig:kinetics}
\end{figure*}

\subsection{Leakage, parity recovery, and poisoning}\label{sec:kinetics}
\label{sec:escape}

Phonons conserve wire parity. They can change the Majorana occupation while bulk quasiparticles are present, but once the bulk empties, the initial Majorana occupation must return if no quasiparticles enter or leave the wire. We use a simplified two-mode model to illustrate this point.

Figure~\ref{fig:kinetics}(a) illustrates the role of pair
correlations using two modes restricted to $(0,0)$ and
$(1,1)$ occupations. The blue solid curve shows the exact occupation
$f_0=f_1=P_{11}$, which also equals the joint occupation
$\langle n_0n_1\rangle$. It obeys
$\dot P_{11}=r_+(1-P_{11})-r_-P_{11}$.
The gray dotted curve shows the product of these exact
mean occupations, $f_0f_1=P_{11}^2$, which is much smaller
than the true joint occupation at low excitation probability.

The orange dashed curve shows the occupation obtained
by solving the factorized kinetic equation,
$\dot f=r_+(1-f)^2-r_-f^2$, with $f_0=f_1=f$.
Replacing the joint occupation by a product underestimates
recombination and causes this approximation to overestimate
the occupation.

For the 0.20-meV pulse, exact leakage and Majorana flips rise together and then decay [Fig.~\ref{fig:kinetics}(b)]. Factorized kinetics instead leaves a spurious residual $f_0\simeq2\Pexc$. It also miscounts Pauli blocking, allowing the bulk partner to enter a Majorana level that is already occupied whenever the pair is present. The initial Majorana occupation recovers for all seven spectra and for either initial occupation, with a mean post-pulse return time of about 130~ns (Table~\ref{tab:pulses}).

Recombination in a large superconductor accelerates with quasiparticle density~\cite{Rothwarf1967,Kaplan1976}. Here, correlated pairs recombine at an almost fixed rate. The exact $\Pflip/u$ curves coincide [Fig.~\ref{fig:kinetics}(c)], and the return time changes little over five decades of energy density. Equilibrium checks and the dependence on phonon energy density are tabulated in Table~\ref{tab:dose}.

Permanent poisoning requires quasiparticle exchange. If a bulk partner escapes before recombination, a Majorana-plus-bulk excitation leaves the Majorana occupation reversed [Fig.~\ref{fig:kinetics}(d)]. The poisoned fraction grows linearly with slow escape and saturates at the Majorana-plus-bulk share of excitations. Without escape, it vanishes; factorized kinetics still predicts roughly two residual flips per excitation.

\begin{figure*}[t]
\includegraphics[width=\textwidth]{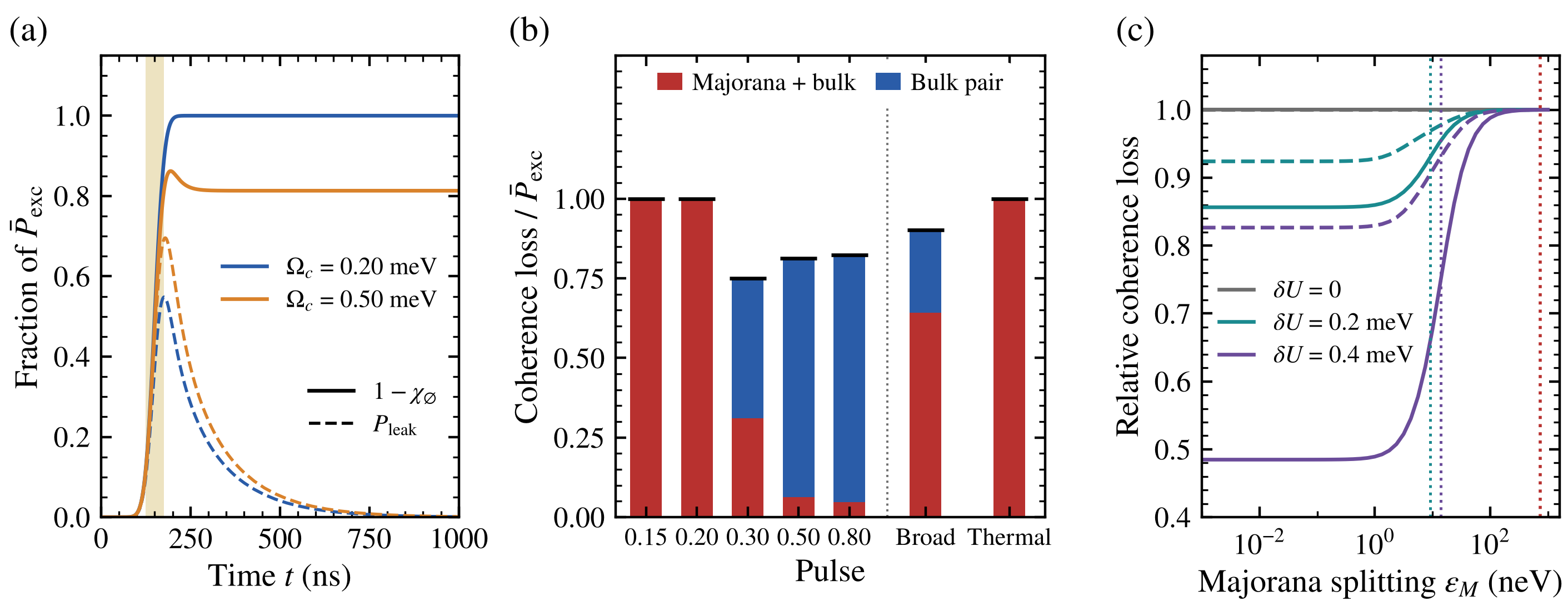}
\caption{Logical coherence and mirror symmetry (rapidly oscillating coherence-transfer terms neglected in a,b). (a) Code-space coherence loss (solid) and leakage from the initially empty-Majorana branch (dashed), normalized to the branch-averaged excitation probability $\Pbar$, for 0.20- and 0.50-meV pulses at $u=0.1$~\evum. Shading marks the pulse FWHM. (b) Final loss per excitation for the seven spectra in Table~\ref{tab:pulses}, resolved into Majorana-plus-bulk and bulk-pair paths involving the Majorana level. (c) Loss versus splitting, varied at fixed wave functions to first order in $u$ and normalized to each wire's loss at $\eps=1~\mu$eV. Solid and dashed curves show 0.15- and 0.20-meV pulses, respectively, for a uniform wire and gate gradients $U(x)=\delta U(x/L-1/2)$. Dotted lines mark $\hbar\Gamma_1/2$ for the gradient wires and the working-point splitting.}
\label{fig:coherence}
\end{figure*}

\subsection{Logical coherence}\label{sec:coherence}

Occupation recovery leaves permanent coherence loss when the exchanged phonons reveal the logical state. Pair creation from an empty Majorana level exchanges energy $E_a+\eps$, whereas promotion from an occupied level exchanges $E_a-\eps$. At the working point, $\hbar/(2\eps)=0.44$~ns is much shorter than the excited-state lifetime, so these phonon energies are distinguishable. For the 0.20-meV pulse, the blue solid curve in
Fig.~\ref{fig:coherence}(a),
$[1-\chi_\varnothing(t)]/\Pbar$, , where
$\Pbar=(P^{(0)}_{\rm exc}+P^{(1)}_{\rm exc})/2$, rises during the pulse
and saturates near unity. The blue dashed leakage curve
subsequently decays to zero, while the coherence loss
remains at this plateau. Thus bulk relaxation restores
the code-space population but leaves a coherence loss
$1-\chi\simeq\Pbar$. Complete dephasing of an excited path corresponds to a logical phase flip with probability $1/2$~\cite{Knapp2018Quantum}.

Bulk-pair creation initially acts identically on both logical states. Coherence survives direct recombination, but is lost if relaxation involves the Majorana level. Most bulk pairs take this second route, giving a final loss of $(0.75\text{--}1.00)\Pbar$ across the seven spectra [Fig.~\ref{fig:coherence}(b), Table~\ref{tab:pulses}]. The transient overshoot in panel (a) comes from coherence carried by excited configurations and later recovered. Independent wires contribute additively at small loss. Escape changes poisoning much more strongly than coherence, since Majorana-plus-bulk paths have already dephased (Table~\ref{tab:coh}).

Reducing $\eps$ removes the energy distinction, but mirror symmetry still makes the two phonon states orthogonal. In a uniform wire, reflection combined with $\sigma_x$ makes $M^B_{a0}(k)M^A_{a0}(k)^*$ odd in $k$, so the cross kernel vanishes (Appendix~\ref{app:phonon-overlap}). The loss consequently persists as $\eps\to0$ [Fig.~\ref{fig:coherence}(c)].

A gate gradient breaks this symmetry and concentrates bulk states near one end, allowing both logical branches to couple to overlapping phonon states (Table~\ref{tab:coh}). Coherence loss then decreases when $2\eps$ falls below the decay width. For the 0.15-meV pulse, the crossover is near $\eps=\hbar\Gamma_1/2\simeq10$~neV, with $\Gamma_1$ the lowest excitation's decay rate. Fig.~\ref{fig:coherence}(c) normalizes each curve to its own loss at $\eps=1~\mu$eV because the gate gradient also changes the excitation probability. At the working point, $2\eps/\hbar$ greatly exceeds the
transition rates, so the cross terms coupling $\chi_S$
to $\chi^*_{S\pm a}$ average out.
\begin{figure*}[t]
\includegraphics[width=\textwidth]{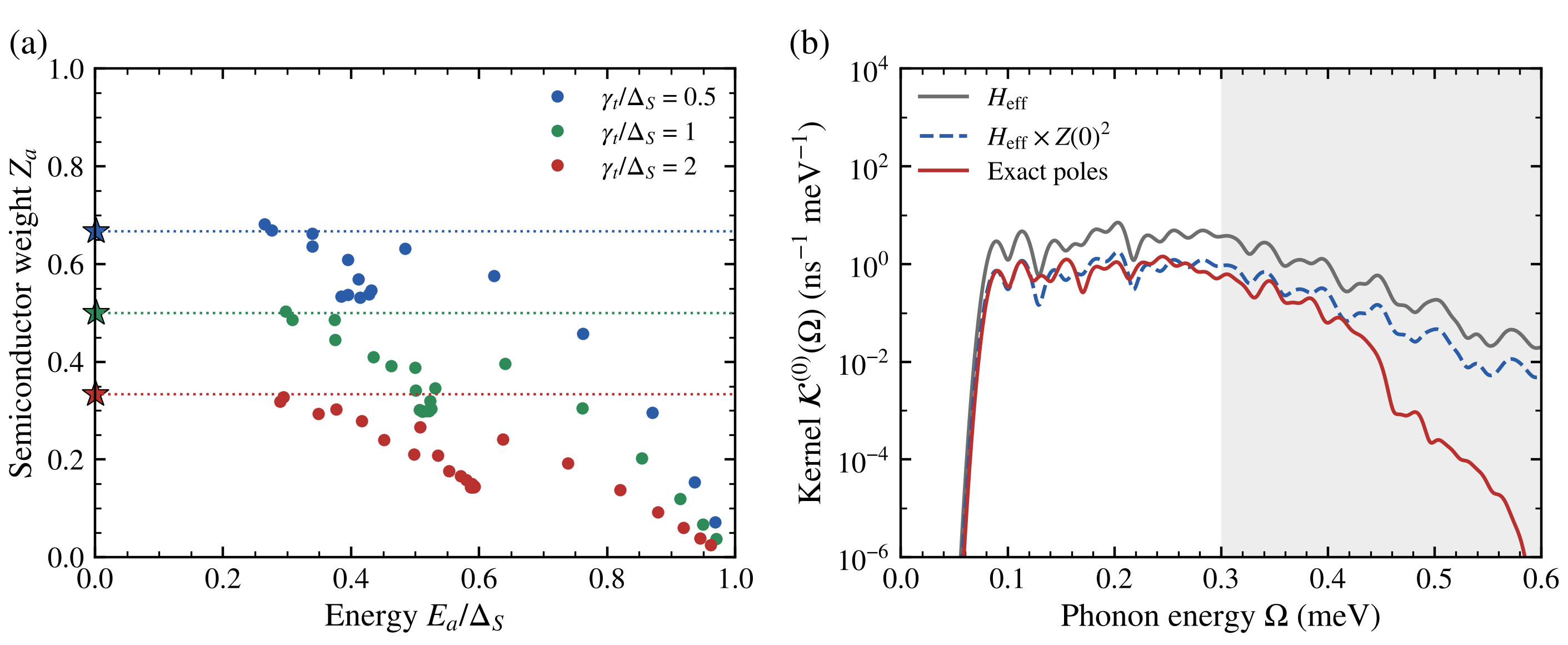}
\caption{Semiconductor weight and its effect on phonon excitation.
The parent superconducting gap is $\Delta_S=0.30$~meV.
(a) Semiconductor weight $Z_a$ versus bound-state energy
for $\gamma_t/\Delta_S=0.5$, 1, and 2. Stars identify the Majorana-derived states; dotted lines
show the low-energy approximation
$Z(0)=(1+\gamma_t/\Delta_S)^{-1}$.
(b) Excitation kernel $\Kp{0}(\Omega)$ for
$\gamma_t=\Delta_S$, calculated using three approaches:
the effective Hamiltonian $H_{\rm eff}$ with
unit-normalized eigenstates (gray);
the same kernel multiplied by $Z(0)^2$ (blue dashed);
and the hybrid bound states, with phonons coupled
to their semiconductor components
$w_a=\sqrt{Z_a}\varphi_a$ (red).
Here $\varphi_a$ is normalized to one.
The shaded region marks energies above the parent gap,
where transitions into the parent continuum become possible, while these transitions are not included.}
\label{fig:parent}
\end{figure*}
\subsection{Hybrid normalization}\label{sec:hybrid}

We compare excitation rates from the effective model with those obtained when the parent superconductor is included explicitly. The ratio $\gamma_t/\Delta_S$ measures the wire--parent tunnel-coupling energy scale relative to the parent gap. Increasing this ratio strengthens hybridization, shifting more of a low-energy state's weight into the parent. Figure~\ref{fig:parent}(a) shows the semiconductor weight $Z_a$ versus state energy for three coupling strengths. Because the phonon coupling considered here acts only in the semiconductor, matrix elements acquire $\sqrt{Z_aZ_b}$ and rates acquire $Z_aZ_b$. At low energies, $Z_a\approx Z(0)=(1+\gamma_t/\Delta_S)^{-1}$.

Figure~\ref{fig:parent}(b) compares three excitation kernels at $\gamma_t=\Delta_S$. The gray curve uses unit-normalized effective states with the bare phonon vertex, assigning their entire weight to the semiconductor. The blue dashed curve corrects the phonon vertex using the constant factor $Z(0)^2=1/4$, while the red curve uses the exact hybrid states and their individual semiconductor weights. Near threshold, hybrid normalization therefore reduces the kernel approximately fourfold, indicating that including hybridization makes phonons less effective at exciting the wire.

\begin{figure*}[t]
\includegraphics[width=\textwidth]{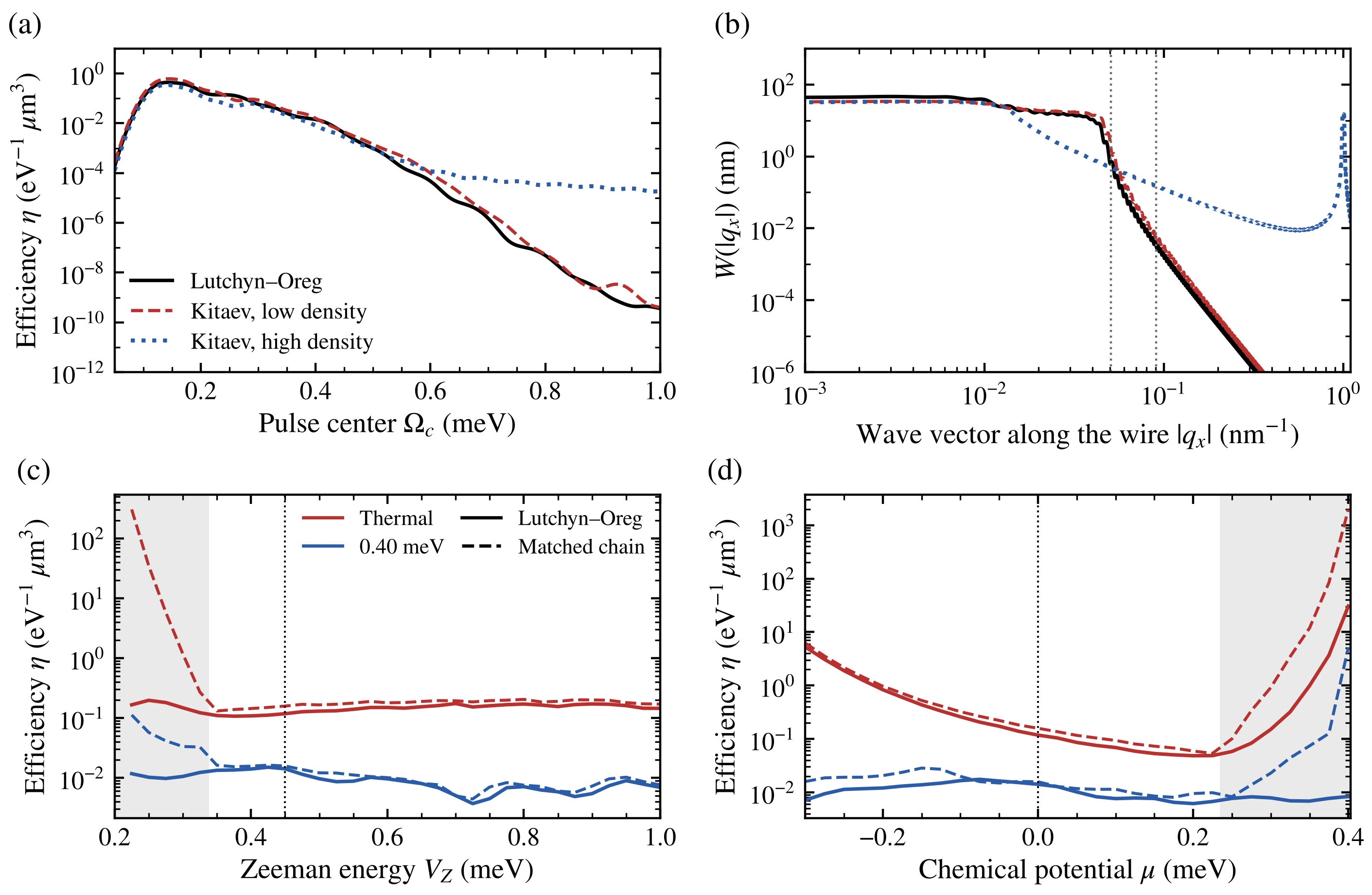}
\caption{Lutchyn--Oreg wire and Kitaev chains matched to its gap and Majorana decay length, at low ($k_0=0.0184$~nm$^{-1}$) and high ($k_0a=1.25$) density. (a) Narrowband excitation efficiency for $\sigma=20~\mu$eV. (b) Fourier weight of Majorana-plus-bulk transition densities below 0.6~meV, normalized to unit area. Dotted lines mark LA and TA wave vectors at threshold. (c),(d) Thermal and 0.40-meV efficiencies versus field and chemical potential, compared with low-density chains matched at each point. Shading marks a gap minimum at $k=0$; dotted lines mark the working point.}
\label{fig:lom}
\end{figure*}

\subsection{Electronic structure effects}\label{sec:lom}

We compare the Lutchyn--Oreg wire with Kitaev chains
matched to its gap and Majorana decay length to identify
which results depend on the semiconductor band structure. The Kitaev chains reproduce parity recovery, the failure of factorized kinetics, and excitation-induced coherence loss, with recovery times within the same order of magnitude of the wire (Table~\ref{tab:lom}). These results follow from parity conservation, pair correlations, and phonons distinguishing the logical states. Uniform Kitaev chains also have vanishing cross kernels by reflection symmetry.

Excitation efficiencies depend more strongly on electronic structure. Below 0.5~meV, the low-density chain agrees with the nanowire within a factor of 1.6, whereas at 1~meV the high-density chain is $5\times10^4$ more efficient than the nanowire [Fig.~\ref{fig:lom}(a)]. Figure~\ref{fig:lom}(b) plots the normalized, symmetrized Fourier weight for Majorana--bulk pair creation, $W(|q_x|) = \frac{S(|q_x|) + S(-|q_x|)}{\int_0^{q_{\max}} [\,S(q) + S(-q)\,]\, dq}$, where $S(k)=\sum_{a\ge1:\,E_a+\epsilon_M\le0.6\,\mathrm{meV}}|M^B_{0a}(k)|^2$, $M^B_{0a}$ is the Fourier transform of the pair-creation transition density defined in Eq.~(11), and $q_{\max}=0.9\pi/a=1.13~\mathrm{nm}^{-1}$. In the low-density chain, the densities governing electron--phonon transitions vary smoothly near the ends, so their Fourier transforms contain little weight at large longitudinal wavevectors. In the high-density chain, sharper spatial variations near the hard-wall boundaries produce substantial high-momentum Fourier weight [Fig.~\ref{fig:lom}(b)]. Energetic acoustic phonons have large wavevectors; those propagating along the chain can therefore couple to these high-momentum components, while coupling to transverse phonons is suppressed by the transverse form factor. Smoothing the high-density chain's transition densities near the ends removes much of this Fourier weight and strongly reduces the excitation efficiency (Table~\ref{tab:lom}).

Matching the gap, decay length, and Fermi momentum reproduces efficiencies within about a factor of two while the gap minimum lies near $k_F$ [Figs.~\ref{fig:lom}(c),(d)]. The Kitaev chain's dependence on $V_Z$ is obtained through parameter matching. At each $V_Z$, we extract the wire's bulk gap $E_g$ and the decay length $\xi$ and wavevector $k_0$ of its slowest-decaying oscillatory zero-energy solution. These three quantities determine $t$, $\mu_K$, and $\Delta_p$ in the Kiteav chain model. Near the topological transition, the wire's lowest-energy bulk excitations occur near $k=0$, while the Majorana mode's spatial oscillations are still determined by states near $k_F$. The matched Kitaev chain cannot reproduce both features simultaneously. It predicts too many low-energy bulk states and overestimate the thermal excitation efficiency by three orders of magnitude (Table~\ref{tab:lom}).

The field-driven and chemical-potential-driven transitions also differ. A small gap makes excitation energetically accessible, but selection rules can still suppress the coupling to phonons. At $k=0$ in the topological phase, the two positive-energy states have the same spin, so spin-conserving density coupling cannot create a pair in them. Near $\mu=0$, the spin and electron--hole structure of the low-energy states makes their phonon-induced pair-creation matrix elements nearly vanish, even at small nonzero momentum. Moving $\mu$ away from zero changes this structure and allows stronger pair creation, even for the same excitation gap. Thus, lowering $V_Z$ to reduce the gap to $25~\mu$eV changes the thermal-pulse efficiency by less than a factor of two, while tuning $\mu$ to reach the same gap increases the efficiency to 31 times its working-point value [Figs.~\ref{fig:lom}(c),(d), Table~\ref{tab:lom}]. Phonon sensitivity therefore depends on the states' spin and particle--hole structure as well as their excitation energies.
\section{Summary and outlook}\label{sec:summary}

Phonon-induced excitations can relax without restoring the quantum information stored in a Majorana qubit. Without quasiparticle exchange with the environment, wire parity is conserved, so the initial Majorana occupation is restored when the bulk empties. The mean post-pulse return time is about 130~ns in the Lutchyn--Oreg wires studied here. Factorized Boltzmann kinetics misses the pair correlations and instead leaves a spurious residual Majorana occupation of about $2\Pexc$ for an initially empty level under a weak pulse. Despite occupation recovery, the exchanged phonons usually distinguish the logical states. For the seven spectra studied, the fractional coherence loss per wire is $0.75$--$1.00$ times its branch-averaged excitation probability. These mechanisms also occur in the matched Kitaev chains. Reducing the Majorana splitting makes the phonons exchanged
by the two logical states closer in energy. In a
mirror-symmetric wire, however, their different spatial
patterns still reveal the logical state. Breaking this
symmetry allows the phonon states to overlap, so a
sufficiently small splitting can reduce coherence loss.

The excitation probability depends strongly on both the phonon spectrum and the electronic structure. At equal phonon energy density and pulse envelope, spectra concentrated near threshold cause far more excitation than higher-energy spectra. Low electron density suppresses coupling to energetic phonons, while a selection rule near the field-driven transition weakens coupling to the modes whose gap closes. Changing the chemical potential weakens this approximate protection away from $k=0$, so wires with the same bulk gap can have very different excitation probabilities. Accounting for the semiconductor weight of hybrid states reduces near-threshold excitation rates by about fourfold relative to the bare-vertex calculation when $\gamma_t=\Delta_S$.

Quantitative device predictions will require band structures and interface tunneling appropriate to the device geometry~\cite{Woods2018}, microscopic electron--phonon couplings~\cite{Ponce2016,Huang2024}, and phonon modes of the wire, shell, and substrate. Coupling within the superconducting shell and transitions into its continuum must also be included~\cite{Huang2024}. Further extensions should address multiband and orbital effects, realistic confinement, the superconducting backbone and island charging energy, correlations between wires exposed to a shared phonon field, and corrections to the Born--Markov treatment. Combining these kinetics with substrate and interface phonon transport, bath back-action, and microscopic escape channels would connect an ionizing event to qubit errors. By linking nonequilibrium phonon spectra to quasiparticle dynamics and the survival of quantum information, this framework provides a microscopic foundation for assessing radiation-induced errors and guiding the design of more resilient Majorana qubits.

\section*{Data availability}
The code and data supporting this article are available from the author upon reasonable request.

\appendix
\section{Model parameters and numerical settings}\label{app:parameters}

The default nanowire parameters are $m^*=0.023\,m_e$, $\alpha=20~\mathrm{meV\,nm}$, $\mu=0$, $V_Z=0.45~\mathrm{meV}$, $\Delta=0.20~\mathrm{meV}$, $L=2~\mu\mathrm{m}$, and $a=2.5$~nm. The two parallel wires are separated by $d=300$~nm and have no tunneling between them. For the phonons, we use the mass density $\rho=5.667~\mathrm{g\,cm^{-3}}$, Voigt-averaged sound velocities $c_L=4.195$ and $c_T=2.352~\mathrm{km\,s^{-1}}$, and LA deformation potential $D=5.08$~eV~\cite{Vurgaftman2001}. The unscreened piezoelectric coupling uses $e_{14}=0.045~\mathrm{C\,m^{-2}}$ and $\varepsilon_r=15.15$~\cite{Madelung2004}. The wire axis is [111], and the Gaussian transverse electron density has radius $R=20$~nm.

The bath is isotropic and incoherent at $T_{\rm bath}=20$~mK. Pulses have a Gaussian time envelope centered at 150~ns with a full width at half maximum of 50~ns. The narrowband spectrum is Gaussian with standard deviation $\sigma=20~\mu\mathrm{eV}$ and center $\Omega_c$; the broadband spectrum has uniform excess occupation from 0.1 to 1.0~meV; and the thermal spectrum is the excess Bose occupation at $T^*=311$~mK over the 20-mK bath. The reference peak excess energy density is $u=0.1~\mathrm{eV}\,(\mu\mathrm{m})^{-3}$, equivalent to 10~keV distributed within a cube of side $46~\mu\mathrm{m}$.

The main kinetic calculation retains 34 quasiparticle modes below 1.2~meV and configurations with at most $k_{\max}=2$ bulk quasiparticles. The scan over phonon energy density uses 17 modes and $k_{\max}=4$. Retaining the coherences of 301 near-degenerate configuration pairs changes the time-integrated leakage by a relative $\le10^{-5}$. Plotted excitation kernels are broadened with a Gaussian of standard deviation $10~\mu\mathrm{eV}$, reduced to $6~\mu\mathrm{eV}$ for the parent-superconductor comparison. That comparison uses $\Delta_S=0.30$~meV, $\gamma_t/\Delta_S=0.5$, 1, and 2, $V_{Z,\rm bare}=2.25\gamma_t$, lattice spacing $a=5$~nm, and bound states with $E<0.97\Delta_S$. The Kitaev-chain parameters $(t,\mu_K,\Delta_p)$ are $(194,0.42,1.37)$~meV for the low-density control and $(9.4,12.9,0.066)$~meV for the high-density control. The latter has $k_0a=1.25$ and $2k_0=1.0~\mathrm{nm^{-1}}$, exceeding the wave vectors of phonons with energies up to 1.5~meV.

\section{Phonon-state overlap and logical coherence}
\label{app:phonon-overlap}

Here we expand the overlap in Eq.~\eqref{eq:phonon-overlap} and distinguishes the loss of coherence along an excitation path from the coherence loss averaged over a pulse. The comparison uses the common phonon energy $\Omega=E_a$; the physical pair-creation and promotion energies are $E_a+\eps$ and $E_a-\eps$, respectively.

\subsection{Direct and cross coefficients}

Set $u=\cos\theta$ and $q_\nu=E_a/(\hbar c_\nu)$. All three coefficients contain the same nonnegative angular weight,
\begin{multline}
w_\nu(u)=\frac{q_\nu}{8\pi^2\hbar\rho c_\nu^2}
\bar C_\nu(\arccos u,q_\nu)\\
\times
\left|F_\perp\!\left(q_\nu\sqrt{1-u^2}\right)\right|^2
\geq0.
\label{eq:overlap-angular-weight}
\end{multline}
Writing out the direct and cross coefficients gives
\begin{align}
\Gamma_a^A&=\sum_\nu\int_{-1}^{1}\!du\,
w_\nu(u)|M^A_{a0}(q_\nu u)|^2,
\label{eq:overlap-direct-A}\\
\Gamma_a^B&=\sum_\nu\int_{-1}^{1}\!du\,
w_\nu(u)|M^B_{a0}(q_\nu u)|^2,
\label{eq:overlap-direct-B}\\
\Gamma_a^\times&=\sum_\nu\int_{-1}^{1}\!du\,
w_\nu(u)M^B_{a0}(q_\nu u)M^A_{a0}(q_\nu u)^*.
\label{eq:overlap-cross-expanded}
\end{align}
The direct coefficients sum squared amplitudes, whereas the cross coefficient sums their complex products. Contributions to the latter can cancel between phonon directions even when both direct coefficients are nonzero.

For $\Gamma_a^A,\Gamma_a^B>0$, define the normalized mode amplitudes
\begin{equation}
f_\nu^X(u)=
\frac{\sqrt{w_\nu(u)}\,M^X_{a0}(q_\nu u)}
{\sqrt{\Gamma_a^X}},
\qquad X=A,B.
\label{eq:overlap-normalized-amplitudes}
\end{equation}
Equations~\eqref{eq:overlap-direct-A} and \eqref{eq:overlap-direct-B} imply
\begin{equation}
\sum_\nu\int_{-1}^{1}\!du\,|f_\nu^X(u)|^2=1.
\label{eq:overlap-amplitude-normalization}
\end{equation}
Substitution into Eq.~\eqref{eq:phonon-overlap} then yields
\begin{equation}
\kappa_a=
\left|\sum_\nu\int_{-1}^{1}\!du\,
f_\nu^A(u)^*f_\nu^B(u)\right|
=|\langle f^A|f^B\rangle|.
\label{eq:overlap-inner-product}
\end{equation}
Thus $\kappa_a$ is the magnitude of a normalized inner product, and $\kappa_a\leq1$ follows from the Cauchy--Schwarz inequality. A zero value means orthogonal mode profiles; unity means identical normalized profiles up to a common phase.

These expressions exclude phonon occupations. The same construction applies to absorption or emission coefficients if $w_\nu(u)$ is replaced everywhere by $w_\nu(u)N_\nu(E_a,t)$ or $w_\nu(u)[N_\nu(E_a,t)+1]$, respectively. The bound requires the same energy and weighting in the two direct coefficients and their cross coefficient.

\subsection{Cancellation from mirror symmetry}

Choose the coordinate origin at the wire center. For the uniform Lutchyn--Oreg wire, reflection combined with $\sigma_x$, denoted by $\mathcal R=\sigma_x\mathcal P_x$, commutes with the Hamiltonian. Eigenstates can therefore be chosen with $\mathcal R\Phi_j=r_j\Phi_j$, where $r_j=\pm1$. Particle--hole conjugation anticommutes with this reflection, so
\begin{equation}
\mathcal R\mathcal C\Phi_0=-r_0\mathcal C\Phi_0.
\label{eq:overlap-reflection-partner}
\end{equation}
Since the density vertex $\tau_z$ is invariant under the reflection, the two transition densities have opposite reflection parity:
\begin{equation}
\begin{aligned}
\rho^A_{a0}(-x)&=r_ar_0\,\rho^A_{a0}(x),\\
\rho^B_{a0}(-x)&=-r_ar_0\,\rho^B_{a0}(x).
\end{aligned}
\label{eq:overlap-density-parities}
\end{equation}
Their Fourier transforms consequently obey
\begin{equation}
\begin{aligned}
M^A_{a0}(-k)&=r_ar_0\,M^A_{a0}(k),\\
M^B_{a0}(-k)&=-r_ar_0\,M^B_{a0}(k).
\end{aligned}
\label{eq:overlap-form-factor-parities}
\end{equation}
The cross product is therefore odd in longitudinal momentum:
\begin{equation}
M^B_{a0}(-k)M^A_{a0}(-k)^*
=-M^B_{a0}(k)M^A_{a0}(k)^*.
\label{eq:overlap-odd-product}
\end{equation}
For the reflection-symmetric phonon weighting used here, $w_\nu(-u)=w_\nu(u)$. The integral in Eq.~\eqref{eq:overlap-cross-expanded} vanishes, giving $\Gamma_a^\times=0$ and $\kappa_a=0$. Breaking mirror symmetry removes this cancellation and permits a nonzero overlap, whose magnitude must be calculated from the resulting wave functions.

The common-energy overlap describes the phonon mode profiles. It does not include the distinguishability of phonons with different energies $E_a\pm\eps$. Consequently, $\kappa_a=1$ alone does not guarantee preservation of coherence: the energy difference and the two branch rates also matter.

\subsection{Orthogonal phonons and pulse-averaged dephasing}

A vanishing $\kappa_a$ removes the coherence-transfer term for the corresponding transition; it does not set the entire logical coherence to zero. Orthogonal phonon states retain information about the logical branch along that excitation path. The effect of a pulse also depends on the probability of taking such a path.

For illustration, suppose that an excursion completely dephases the qubit with the same probability $P_{\rm d}$ on both logical branches, while the remaining events leave it unchanged. After recovery,
\begin{equation}
\begin{aligned}
\rho_{01}^{\,f}&=(1-P_{\rm d})\rho_{01}^{\,i},\\
D&=1-\frac{|\rho_{01}^{\,f}|}{|\rho_{01}^{\,i}|}
=P_{\rm d}.
\end{aligned}
\label{eq:overlap-pulse-averaged-loss}
\end{equation}
Even complete dephasing along every excited path therefore produces only a small average loss when those paths are rare. For example, $P_{\rm d}=0.01$ gives a 1\% coherence loss. The excitation probability determines how often the wire leaves its empty-bulk state, whereas $\kappa_a$ helps determine whether particular transitions preserve the coherence carried by those excursions. Bulk pairs that recombine without involving the Majorana level can return coherence to the code space. In general, the final loss must therefore be obtained from the coupled $\chi_S$ dynamics rather than from $\kappa_a$ or $\Pexc$ alone.

\section{Coefficient matrices and coherence evolution}
\label{app:coefficient-matrix}

The two-channel matrices in Eq.~\eqref{eq:paired-rate-matrices} act on transition operators. Their off-diagonal entries describe interference between the pair-creation and promotion amplitudes. This appendix gives their explicit structure and the conditions under which they generate completely positive evolution.

\subsection{Direct and cross terms}

Before including phonon occupations, the common-energy matrix in the ordered channel basis $(B,A)$ is
\begin{equation}
\mathsf C_a^{(0)}=
\begin{pmatrix}
\Gamma_a^B&\Gamma_a^\times\\
(\Gamma_a^\times)^*&\Gamma_a^A
\end{pmatrix}.
\label{eq:coefficient-bare-matrix}
\end{equation}
The operators are $F_{aB}^+=c_a^\dagger c_0^\dagger$ and $F_{aA}^+=c_a^\dagger c_0$, with reverse operators $F_{aX}^-=(F_{aX}^+)^\dagger$. For example, the absorption contribution is
\begin{multline}
\mathcal D_a^+[\rho]=
\sum_{X,Y\in\{B,A\}}(\mathsf C_a^+)_{XY}\\
\times\left(F_{aX}^+\rho F_{aY}^{+\dagger}
-\frac12\{F_{aY}^{+\dagger}F_{aX}^+,\rho\}\right).
\label{eq:coefficient-channel-evolution}
\end{multline}
The terms $X=Y$ are the ordinary pair-creation and promotion channels; $X\ne Y$ transfer coherence between the corresponding configurations.

Write $\mathsf C_{a\nu}^{(0)}$ for the branch-$\nu$ contribution to Eq.~\eqref{eq:coefficient-bare-matrix}, with every entry evaluated at $E_a$. A consistent common-energy construction including phonon occupations is
\begin{equation}
\begin{aligned}
\widetilde{\mathsf C}_a^+(t)
&=\sum_\nu N_\nu(E_a,t)\mathsf C_{a\nu}^{(0)},\\
\widetilde{\mathsf C}_a^-(t)
&=\sum_\nu[N_\nu(E_a,t)+1]\mathsf C_{a\nu}^{(0)*}.
\end{aligned}
\label{eq:coefficient-common-energy-matrices}
\end{equation}
Complex conjugation in the second line follows because emission uses the conjugate microscopic amplitudes and the adjoint transition operators. The tilde distinguishes this common-energy construction from a matrix whose diagonal entries retain the physical transition energies $E_a\pm\eps$.

\subsection{Positivity and its energy convention}

For the matrix in Eq.~\eqref{eq:coefficient-bare-matrix}, the diagonal entries are nonnegative and
\begin{equation}
\begin{aligned}
\det\mathsf C_a^{(0)}
&=\Gamma_a^A\Gamma_a^B-|\Gamma_a^\times|^2\\
&=\Gamma_a^A\Gamma_a^B(1-\kappa_a^2)\geq0.
\end{aligned}
\label{eq:coefficient-determinant}
\end{equation}
Together these conditions make the Hermitian $2\times2$ matrix positive semidefinite. Equivalently, its eigenvalues are
\begin{multline}
\lambda_\pm=\frac12\left[\Gamma_a^A+\Gamma_a^B\right.\\
\left.\pm\sqrt{(\Gamma_a^A-\Gamma_a^B)^2
+4|\Gamma_a^\times|^2}\right]\geq0.
\label{eq:coefficient-eigenvalues}
\end{multline}
The same argument holds for each phonon branch separately. Multiplication by the nonnegative Bose factors and summation over branches therefore preserve positivity in Eq.~\eqref{eq:coefficient-common-energy-matrices}.

More generally, for the matrices with diagonal rates $d_{aX}^\pm$ and cross coefficients $r_a^\pm$ in Eq.~\eqref{eq:paired-rate-matrices}, positivity holds if and only if
\begin{equation}
d_{aA}^\pm\geq0,\qquad d_{aB}^\pm\geq0,\qquad
|r_a^\pm|^2\leq d_{aA}^\pm d_{aB}^\pm.
\label{eq:coefficient-physical-rate-bound}
\end{equation}
The bare bound $\kappa_a\leq1$ proves this inequality only when the direct and cross terms use the same energy and phonon weighting. If the diagonal rates instead use $E_a-\eps$ and $E_a+\eps$, while the cross term is evaluated at $E_a$, Eq.~\eqref{eq:coefficient-physical-rate-bound} is an additional condition that must be checked; it does not follow from the common-energy overlap alone. A consistently positive common-energy block is provided by Eq.~\eqref{eq:coefficient-common-energy-matrices}. When rapid oscillations suppress the cross terms, the separate physical rates give nonnegative diagonal matrices.

For any positive block, diagonalization $\mathsf C=U\,\mathrm{diag}(\lambda_r)U^\dagger$ gives jump operators
\begin{equation}
L_r=\sqrt{\lambda_r}\sum_X U_{Xr}F_X.
\label{eq:coefficient-diagonalized-jumps}
\end{equation}
Its contribution becomes
$\sum_r[L_r\rho L_r^\dagger-\{L_r^\dagger L_r,\rho\}/2]$.
The resulting time-local evolution is trace preserving and completely positive when all retained blocks are positive. This statement applies before configuration truncation, or to a space that includes the truncation sink. Probability discarded beyond the cutoff makes the retained block trace decreasing; the outgoing rates must still be included in its diagonal loss terms. A zero cross coefficient leaves two independent transition channels; it does not remove their direct rates. A rank-one positive block has one combined jump operator.

\subsection{Configuration elements, phases, and the spectator wire}

The component equation~\eqref{eq:coherence-component-master} follows by taking matrix elements of Eq.~\eqref{eq:coherence-operator-master}. Each allowed jump changes definite occupations. Moreover, the products $F_{aA}^{+\dagger}F_{aB}^+$ and $F_{aB}^{+\dagger}F_{aA}^+$ vanish, as do the corresponding emission products. Consequently, the anticommutator contains a diagonal loss operator and contributes
$-(\Lambda_n+\Lambda_m)\rho_{nm}/2$.
The jump term gives $K_{nm;kl}\rho_{kl}$, including the signs generated by the fermionic ordering in Eq.~\eqref{eq:occupation-configuration}. This construction also applies to escape jumps and avoids assigning their coherence-transfer signs from population rates alone.

The spectator occupation remains part of the full two-wire state label. Without escape, even total parity fixes the explicit version of the shorthand coherence in Eq.~\eqref{eq:cS}:
\begin{equation}
\chi_S=
\frac{\langle0,S;r_S|\rho|1,S;1-r_S\rangle}{C_0},
\qquad
r_S=|S|\bmod2,
\label{eq:coefficient-explicit-spectator}
\end{equation}
where the label after the semicolon is the partner wire's Majorana occupation and $C_0=\rho_{01}(0)\ne0$. At $S=\varnothing$, this is the original logical coherence between $|00\rangle$ and $|11\rangle$. It is not an off-diagonal element of the single-wire reduced density matrix. With escape, the total wire parity can change, so the partner occupation must be retained independently rather than inferred from $r_S$.

Finally, Hermiticity relates unnormalized density-matrix elements by $\rho_{mn}=\rho_{nm}^*$. For normalized coherences and a general complex $C_0$, this becomes
\begin{equation}
\frac{\rho_{mn}}{C_0}
=\frac{C_0^*}{C_0}
\left(\frac{\rho_{nm}}{C_0}\right)^*.
\label{eq:coefficient-normalized-conjugation}
\end{equation}
A transfer term written only with $\chi_S^*$ therefore assumes real $C_0$ or must retain the displayed phase factor. The component equation for $\rho_{nm}$ is valid for either convention; the normalized coherences are formed by division by $C_0$.

\bibliography{refs}

@article{Kitaev2001,
  author  = {Kitaev, A. Yu.},
  title   = {Unpaired {Majorana} fermions in quantum wires},
  journal = {Phys.-Usp.},
  volume  = {44},
  number  = {10S},
  pages   = {131},
  year    = {2001},
  doi     = {10.1070/1063-7869/44/10S/S29}
}

@article{Lutchyn2010,
  author  = {Lutchyn, Roman M. and Sau, Jay D. and Das Sarma, S.},
  title   = {{Majorana} Fermions and a Topological Phase Transition in Semiconductor-Superconductor Heterostructures},
  journal = {Phys. Rev. Lett.},
  volume  = {105},
  pages   = {077001},
  year    = {2010},
  doi     = {10.1103/PhysRevLett.105.077001}
}

@article{Oreg2010,
  author  = {Oreg, Yuval and Refael, Gil and von Oppen, Felix},
  title   = {Helical Liquids and {Majorana} Bound States in Quantum Wires},
  journal = {Phys. Rev. Lett.},
  volume  = {105},
  pages   = {177002},
  year    = {2010},
  doi     = {10.1103/PhysRevLett.105.177002}
}

@article{Alicea2012,
  author  = {Alicea, Jason},
  title   = {New directions in the pursuit of {Majorana} fermions in solid state systems},
  journal = {Rep. Prog. Phys.},
  volume  = {75},
  pages   = {076501},
  year    = {2012},
  doi     = {10.1088/0034-4885/75/7/076501}
}

@article{Lutchyn2018,
  author  = {Lutchyn, R. M. and Bakkers, E. P. A. M. and Kouwenhoven, L. P. and Krogstrup, P. and Marcus, C. M. and Oreg, Y.},
  title   = {{Majorana} zero modes in superconductor-semiconductor heterostructures},
  journal = {Nat. Rev. Mater.},
  volume  = {3},
  pages   = {52},
  year    = {2018},
  doi     = {10.1038/s41578-018-0003-1}
}

@article{Aghaee2023,
  author        = {Aghaee, Morteza and others},
  collaboration = {Microsoft Quantum},
  title         = {{InAs-Al} hybrid devices passing the topological gap protocol},
  journal       = {Phys. Rev. B},
  volume        = {107},
  pages         = {245423},
  year          = {2023},
  doi           = {10.1103/PhysRevB.107.245423}
}

@article{Aghaee2025,
  author        = {Aghaee, Morteza and others},
  collaboration = {Microsoft Azure Quantum},
  title         = {Interferometric single-shot parity measurement in {InAs--Al} hybrid devices},
  journal       = {Nature},
  volume        = {638},
  pages         = {651},
  year          = {2025},
  doi           = {10.1038/s41586-024-08445-2}
}

@article{Rainis2012,
  author  = {Rainis, Diego and Loss, Daniel},
  title   = {{Majorana} qubit decoherence by quasiparticle poisoning},
  journal = {Phys. Rev. B},
  volume  = {85},
  pages   = {174533},
  year    = {2012},
  doi     = {10.1103/PhysRevB.85.174533}
}

@article{Karzig2021,
  author  = {Karzig, Torsten and Cole, William S. and Pikulin, Dmitry I.},
  title   = {Quasiparticle Poisoning of {Majorana} Qubits},
  journal = {Phys. Rev. Lett.},
  volume  = {126},
  pages   = {057702},
  year    = {2021},
  doi     = {10.1103/PhysRevLett.126.057702}
}

@article{Wilen2021,
  author  = {Wilen, C. D. and others},
  title   = {Correlated charge noise and relaxation errors in superconducting qubits},
  journal = {Nature},
  volume  = {594},
  pages   = {369},
  year    = {2021},
  doi     = {10.1038/s41586-021-03557-5}
}

@article{McEwen2022,
  author  = {McEwen, Matt and others},
  title   = {Resolving catastrophic error bursts from cosmic rays in large arrays of superconducting qubits},
  journal = {Nat. Phys.},
  volume  = {18},
  pages   = {107},
  year    = {2022},
  doi     = {10.1038/s41567-021-01432-8}
}

@article{Martinis2021,
  author  = {Martinis, John M.},
  title   = {Saving superconducting quantum processors from decay and correlated errors generated by gamma and cosmic rays},
  journal = {npj Quantum Inf.},
  volume  = {7},
  pages   = {90},
  year    = {2021},
  doi     = {10.1038/s41534-021-00431-0}
}

@article{Huang2024,
  author  = {Huang, Han and Kielar, Samuel and Tian, Zhiting},
  title   = {Ab initio calculation of nonequilibrium quasiparticle-phonon dynamics in superconductors},
  journal = {AVS Quantum Sci.},
  volume  = {6},
  pages   = {016201},
  year    = {2024},
  doi     = {10.1116/5.0188992}
}

@article{Aseev2019,
  author  = {Aseev, Pavel P. and Marra, Pasquale and Stano, Peter and Klinovaja, Jelena and Loss, Daniel},
  title   = {Degeneracy lifting of {Majorana} bound states due to electron-phonon interactions},
  journal = {Phys. Rev. B},
  volume  = {99},
  pages   = {205435},
  year    = {2019},
  doi     = {10.1103/PhysRevB.99.205435}
}

@article{Song2018,
  author  = {Song, Yang and Das Sarma, S.},
  title   = {Phonon-induced {Majorana} qubit relaxation in tunnel-coupled two-island topological superconductors},
  journal = {Phys. Rev. B},
  volume  = {98},
  pages   = {075159},
  year    = {2018},
  doi     = {10.1103/PhysRevB.98.075159}
}

@article{Knapp2018,
  author  = {Knapp, Christina and Karzig, Torsten and Lutchyn, Roman M. and Nayak, Chetan},
  title   = {Dephasing of {Majorana}-based qubits},
  journal = {Phys. Rev. B},
  volume  = {97},
  pages   = {125404},
  year    = {2018},
  doi     = {10.1103/PhysRevB.97.125404}
}

@article{Kaestle2020,
  author  = {Kaestle, Oliver and Sun, Yue and Hu, Ying and Carmele, Alexander},
  title   = {Memory-critical dynamical buildup of phonon-dressed {Majorana} fermions},
  journal = {Phys. Rev. B},
  volume  = {102},
  pages   = {245115},
  year    = {2020},
  doi     = {10.1103/PhysRevB.102.245115}
}

@article{Yelton2024,
  author  = {Yelton, Eric and Larson, Clayton P. and Iaia, Vito and Dodge, Kenneth and La Magna, Guglielmo and Baity, Paul G. and Pechenezhskiy, Ivan V. and McDermott, Robert and Kurinsky, Noah and Catelani, Gianluigi and Plourde, Britton L. T.},
  title   = {Modeling phonon-mediated quasiparticle poisoning in superconducting qubit arrays},
  journal = {Phys. Rev. B},
  volume  = {110},
  pages   = {024519},
  year    = {2024},
  doi     = {10.1103/PhysRevB.110.024519}
}

@article{Rothwarf1967,
  author  = {Rothwarf, Allen and Taylor, B. N.},
  title   = {Measurement of Recombination Lifetimes in Superconductors},
  journal = {Phys. Rev. Lett.},
  volume  = {19},
  pages   = {27},
  year    = {1967},
  doi     = {10.1103/PhysRevLett.19.27}
}

@article{Kaplan1976,
  author  = {Kaplan, S. B. and Chi, C. C. and Langenberg, D. N. and Chang, J. J. and Jafarey, S. and Scalapino, D. J.},
  title   = {Quasiparticle and phonon lifetimes in superconductors},
  journal = {Phys. Rev. B},
  volume  = {14},
  pages   = {4854},
  year    = {1976},
  doi     = {10.1103/PhysRevB.14.4854}
}

@article{WilsonProber2004,
  author  = {Wilson, C. M. and Prober, D. E.},
  title   = {Quasiparticle number fluctuations in superconductors},
  journal = {Phys. Rev. B},
  volume  = {69},
  pages   = {094524},
  year    = {2004},
  doi     = {10.1103/PhysRevB.69.094524}
}

@article{Maisi2013,
  author  = {Maisi, V. F. and Lotkhov, S. V. and Kemppinen, A. and Heimes, A. and Muhonen, J. T. and Pekola, J. P.},
  title   = {Excitation of single quasiparticles in a small superconducting {Al} island connected to normal-metal leads by tunnel junctions},
  journal = {Phys. Rev. Lett.},
  volume  = {111},
  pages   = {147001},
  year    = {2013},
  doi     = {10.1103/PhysRevLett.111.147001}
}

@article{Olivares2014,
  author  = {Olivares, D. G. and Levy Yeyati, A. and Bretheau, L. and Girit, {\c{C}}. {\"O}. and Pothier, H. and Urbina, C.},
  title   = {Dynamics of quasiparticle trapping in {Andreev} levels},
  journal = {Phys. Rev. B},
  volume  = {89},
  pages   = {104504},
  year    = {2014},
  doi     = {10.1103/PhysRevB.89.104504}
}

@article{Pedrocchi2015,
  author  = {Pedrocchi, Fabio L. and Bonesteel, N. E. and DiVincenzo, David P.},
  title   = {{Monte Carlo} studies of the properties of the {Majorana} quantum error correction code: Is self-correction possible during braiding?},
  journal = {Phys. Rev. B},
  volume  = {92},
  pages   = {115441},
  year    = {2015},
  doi     = {10.1103/PhysRevB.92.115441}
}

@article{PedrocchiDiVincenzo2015,
  author  = {Pedrocchi, Fabio L. and DiVincenzo, David P.},
  title   = {{Majorana} Braiding with Thermal Noise},
  journal = {Phys. Rev. Lett.},
  volume  = {115},
  pages   = {120402},
  year    = {2015},
  doi     = {10.1103/PhysRevLett.115.120402}
}

@article{Ippoliti2016,
  author  = {Ippoliti, Matteo and Rizzi, Matteo and Giovannetti, Vittorio and Mazza, Leonardo},
  title   = {Quantum memories with zero-energy {Majorana} modes and experimental constraints},
  journal = {Phys. Rev. A},
  volume  = {93},
  pages   = {062325},
  year    = {2016},
  doi     = {10.1103/PhysRevA.93.062325}
}

@article{Breckwoldt2022,
  author  = {Breckwoldt, Niels and Posske, Thore and Thorwart, Michael},
  title   = {Bath-induced decoherence in finite-size {Majorana} wires at non-zero temperature},
  journal = {New J. Phys.},
  volume  = {24},
  pages   = {013033},
  year    = {2022},
  doi     = {10.1088/1367-2630/ac46e2}
}

@article{Schmidt2012,
  author  = {Schmidt, Manuel J. and Rainis, Diego and Loss, Daniel},
  title   = {Decoherence of {Majorana} qubits by noisy gates},
  journal = {Phys. Rev. B},
  volume  = {86},
  pages   = {085414},
  year    = {2012},
  doi     = {10.1103/PhysRevB.86.085414}
}

@article{Cheng2012,
  author  = {Cheng, Meng and Lutchyn, Roman M. and Das Sarma, S.},
  title   = {Topological protection of {Majorana} qubits},
  journal = {Phys. Rev. B},
  volume  = {85},
  pages   = {165124},
  year    = {2012},
  doi     = {10.1103/PhysRevB.85.165124}
}

@misc{Alase2025,
  author        = {Alase, Abhijeet and Goffage, Marcus C. and Cassidy, Maja C. and Coppersmith, Susan N.},
  title         = {Decoherence of {Majorana} qubits by $1/f$ noise},
  year          = {2025},
  eprint        = {2506.22394},
  archivePrefix = {arXiv},
  primaryClass  = {cond-mat.mes-hall}
}

@article{Knapp2018Quantum,
  author  = {Knapp, Christina and Beverland, Michael and Pikulin, Dmitry I. and Karzig, Torsten},
  title   = {Modeling noise and error correction for {Majorana}-based quantum computing},
  journal = {Quantum},
  volume  = {2},
  pages   = {88},
  year    = {2018},
  doi     = {10.22331/q-2018-09-03-88}
}

@article{Mazza2013,
  author  = {Mazza, Leonardo and Rizzi, Matteo and Lukin, Mikhail D. and Cirac, J. Ignacio},
  title   = {Robustness of quantum memories based on {Majorana} zero modes},
  journal = {Phys. Rev. B},
  volume  = {88},
  pages   = {205142},
  year    = {2013},
  doi     = {10.1103/PhysRevB.88.205142}
}

@article{Mishmash2020,
  author  = {Mishmash, Ryan V. and Bauer, Bela and von Oppen, Felix and Alicea, Jason},
  title   = {Dephasing and leakage dynamics of noisy {Majorana}-based qubits: Topological versus {Andreev}},
  journal = {Phys. Rev. B},
  volume  = {101},
  pages   = {075404},
  year    = {2020},
  doi     = {10.1103/PhysRevB.101.075404}
}

@article{Konschelle2013,
  author  = {Konschelle, Fran{\c{c}}ois and Hassler, Fabian},
  title   = {Effects of nonequilibrium noise on a quantum memory encoded in {Majorana} zero modes},
  journal = {Phys. Rev. B},
  volume  = {88},
  pages   = {075431},
  year    = {2013},
  doi     = {10.1103/PhysRevB.88.075431}
}

@article{Munk2019,
  author  = {Munk, Morten I. K. and Egger, Reinhold and Flensberg, Karsten},
  title   = {Fidelity and visibility loss in {Majorana} qubits by entanglement with environmental modes},
  journal = {Phys. Rev. B},
  volume  = {99},
  pages   = {155419},
  year    = {2019},
  doi     = {10.1103/PhysRevB.99.155419}
}

@article{Akhmerov2010,
  author  = {Akhmerov, A. R.},
  title   = {Topological quantum computation away from the ground state using {Majorana} fermions},
  journal = {Phys. Rev. B},
  volume  = {82},
  pages   = {020509(R)},
  year    = {2010},
  doi     = {10.1103/PhysRevB.82.020509}
}

@article{Stanescu2017,
  author  = {Stanescu, Tudor D. and Das Sarma, Sankar},
  title   = {Proximity-induced low-energy renormalization in hybrid semiconductor-superconductor {Majorana} structures},
  journal = {Phys. Rev. B},
  volume  = {96},
  pages   = {014510},
  year    = {2017},
  doi     = {10.1103/PhysRevB.96.014510}
}

@book{Breuer2002,
  author    = {Breuer, Heinz-Peter and Petruccione, Francesco},
  title     = {The Theory of Open Quantum Systems},
  publisher = {Oxford University Press},
  address   = {Oxford},
  year      = {2002}
}

@article{Vurgaftman2001,
  author  = {Vurgaftman, I. and Meyer, J. R. and Ram-Mohan, L. R.},
  title   = {Band parameters for {III-V} compound semiconductors and their alloys},
  journal = {J. Appl. Phys.},
  volume  = {89},
  pages   = {5815},
  year    = {2001},
  doi     = {10.1063/1.1368156}
}

@book{Madelung2004,
  author    = {Madelung, Otfried},
  title     = {Semiconductors: Data Handbook},
  edition   = {3rd},
  publisher = {Springer},
  address   = {Berlin},
  year      = {2004},
  doi       = {10.1007/978-3-642-18865-7}
}

@article{Woods2018,
  author  = {Woods, Benjamin D. and Stanescu, Tudor D. and Das Sarma, Sankar},
  title   = {Effective theory approach to the {Schr{\"o}dinger-Poisson} problem in semiconductor {Majorana} devices},
  journal = {Phys. Rev. B},
  volume  = {98},
  pages   = {035428},
  year    = {2018},
  doi     = {10.1103/PhysRevB.98.035428}
}

@article{Ponce2016,
  author  = {Ponc{\'e}, S. and Margine, E. R. and Verdi, C. and Giustino, F.},
  title   = {{EPW}: Electron-phonon coupling, transport and superconducting properties using maximally localized {Wannier} functions},
  journal = {Comput. Phys. Commun.},
  volume  = {209},
  pages   = {116},
  year    = {2016},
  doi     = {10.1016/j.cpc.2016.07.028}
}

\clearpage
\onecolumngrid
\section*{Supplemental Information}
\renewcommand{\arraystretch}{1.08}
\setcounter{table}{0}
\renewcommand{\thetable}{S\arabic{table}}
\providecommand{\theHtable}{}
\renewcommand{\theHtable}{SI.\arabic{table}}

Unless noted, values use the working-point wire, $T=20$~mK, $L=2~\mu$m, and $u=0.1$~\evum.
Excitation data use the initially empty Majorana level ($p=0$).
Excitation rates are per unit phonon occupation; efficiencies are in \effu.
Semicolon-separated values follow the order of the listed conditions.

\begin{table}[!htbp]
\centering
\small
\caption{Spectrum and excitation. PE denotes piezoelectric coupling.}
\label{tab:scales}
\begin{tabular}{@{}p{0.35\textwidth}p{0.29\textwidth}p{0.30\textwidth}@{}}
\toprule
Quantity & Condition & Value \\
\midrule
\multicolumn{3}{@{}l}{\textbf{Spectrum}}\\[2pt]
$E_g$; $E_1$ & $E_1-E_g\approx13~\mu$eV & 0.1260; 0.1394~meV \\
$\eps$ & & 0.746~$\mu$eV \\
$k_F$; $\pi/k_F$ & Lower helical band & 0.019~nm$^{-1}$; 167~nm \\
Majorana-density period & & 170~nm \\
$\xi$; $L/\xi$ & & 355~nm; 5.6 \\
Normal-state density; wavelength; velocity & $k_F/\pi$; $2\pi/k_F$; $v_F$ & $\approx6~\mu$m$^{-1}$; $\approx330$~nm; $\approx70$~km\,s$^{-1}$ \\
Relative level error & $a=2.5$ vs 1.25~nm; checked to 10~nm & $\le2\times10^{-4}$ \\
\midrule
\multicolumn{3}{@{}l}{\textbf{Excitation kernel}}\\[2pt]
First pair: energy; rate & $E_1+\eps$ & 0.1401~meV; $2.35\times10^7$~s$^{-1}$ \\
First bulk-pair energy & $E_1+E_2$ & 0.287~meV \\
Total rate & All excitation lines & $2.72\times10^8$~s$^{-1}$ \\
Rate fraction & 0.14--0.30; $>0.5$~meV & 67\%; 0.9\% \\
Promotion vs pair-creation rates & Corresponding lines & $-10\%$ to $+26\%$ \\
Fourier weight & $|q_x|\lesssim0.04$~nm$^{-1}$; pair energies $<0.6$~meV & 90\% \\
Phonon wave vector & 0.14~meV: LA; TA & 0.051; 0.090~nm$^{-1}$ \\
Angular rate fraction & $\theta>70^\circ$; $>60^\circ$ & 87\%; 95\% \\
Transverse cutoff & $\hbar c_\nu/R$: LA; TA & 0.14; 0.077~meV \\
Sound-speed; wave-vector ratios & $(c_L/c_T)^3$; $q_{\rm TA}/q_{\rm LA}$ (equal energy) & 5.7; 1.8 \\
Rate fraction by coupling & Deformation; LA PE; TA PE & 77\%; 15\%; 8\% \\
\midrule
\multicolumn{3}{@{}l}{\textbf{Pulse efficiency}}\\[2pt]
$\eta$, narrowband & $\Omega_c=0.142$; 0.40~meV & 0.44; $1.4\times10^{-2}$ \\
 & $\Omega_c=0.50$; 1.0~meV & $9.8\times10^{-4}$; $3.6\times10^{-10}$ \\
 & $\sigma=5~\mu$eV, strongest line & $>1$ \\
$\eta$, broadband & 94\% of energy above 0.5~meV & $1.95\times10^{-3}$ \\
$\eta$, thermal & $T^*=311$~mK & 0.118 \\
\midrule
\multicolumn{3}{@{}l}{\textbf{Field, chemical potential, and radius}}\\[2pt]
Gate-driven transition & $V_Z=0.45$~meV & $\mu=0.40$~meV \\
Excitation threshold & Away from transitions & 0.08--0.16~meV; $\le E_g+0.06$~meV \\
Total rate & Field scan; gate scan & (1.6--3.7); (1.6--9.8)$\times10^8$~s$^{-1}$ \\
Kernel cutoff & & $\approx0.8$~meV \\
Gap minimum at $k=0$ & Field scan; gate scan & $V_Z<0.338$~meV; $\mu>0.234$~meV \\
Total rate & $R=10$; 20; 40~nm & $4.0\times10^9$; $2.7\times10^8$; $1.1\times10^7$~s$^{-1}$ \\
Radius dependence & Total rate; first-line variation, $R=10$--40~nm & $\propto R^{-4.3}$; 13-fold \\
PE fraction at threshold & [111], [110]; [100] & 0.58--0.60; 0.53 \\
TA fraction of PE & [110]; other studied axes & $\approx1/2$; majority \\
PE fraction $<10\%$ & Phonon energy & $>0.25$--0.39~meV \\
\bottomrule
\end{tabular}
\end{table}

\clearpage
\begin{table}[!htbp]
\centering
\small
\caption{Parity, pulse strength, and escape. Energy-density and escape scans use broadband pulses spanning 0.1--0.6 and 0.1--1.0~meV, respectively.}
\label{tab:dose}
\begin{tabular}{@{}p{0.35\textwidth}p{0.29\textwidth}p{0.30\textwidth}@{}}
\toprule
Quantity & Condition & Value \\
\midrule
\multicolumn{3}{@{}l}{\textbf{Parity-projected equilibrium}}\\[2pt]
Majorana occupation $\approx1$ & Odd wire parity & $T\lesssim0.3$~K \\
$\eps/k_B$ & & 8.7~mK \\
\midrule
\multicolumn{3}{@{}l}{\textbf{Phonon energy density $u$ (\evum)}}\\[2pt]
$\Pexc/u$ & $u\to0$ & $1.50\times10^{-2}$~\effu \\
Reduction of $\Pexc/u$ & $u=1$; 10; 100 & 0.7\%; 7\%; 48\% \\
$\tret$, initially empty level & $u=0.001$--100 & 127.1--128.5~ns \\
$\tret$, initially occupied level & $u=0.001$--10 & 136.3--136.5~ns \\
Factorized residual $f_0/\Pexc$ & $u\to0$; $u=100$ & 2.00; 0.64 \\
Discarded weight & $k_{\max}=4$, $u=10$; 100 & $6\times10^{-6}$; $7\times10^{-3}$ \\
Peak-leakage change & $k_{\max}=4$ vs 2, $u=10$ & 0.3\% \\
\midrule
\multicolumn{3}{@{}l}{\textbf{Escape: residual flips per excitation}}\\[2pt]
Exact & Slow escape & $\Gesc\times99$~ns \\
 & $\Gesc=10^4$; $10^7$~s$^{-1}$ & $9.9\times10^{-4}$; 0.42 \\
Half-saturation escape rate & & $\approx6\times10^6$~s$^{-1}$ \\
Saturation value & & 0.642 \\
Majorana-plus-bulk excitation fraction & & 0.641 \\
Factorized & Slow escape & $\approx2$ \\
Factorized/exact & $\Gesc=10^4$~s$^{-1}$ & $2\times10^3$ \\
\bottomrule
\end{tabular}
\par\smallskip
\parbox{0.94\textwidth}{\footnotesize The reduction of $\Pexc/u$ follows saturation of $\Pexc=1-e^{-x}$. Escape is evaluated at the reference $u$.}
\end{table}

\clearpage
\begin{table}[!htbp]
\centering
\small
\caption{Seven pulse spectra: excitation, recovery, and coherence.}
\label{tab:pulses}
\setlength{\tabcolsep}{5pt}
\renewcommand{\arraystretch}{1.12}
\begin{tabular*}{\textwidth}{@{\extracolsep{\fill}}lrrrrr@{}}
\toprule
\multicolumn{6}{@{}l}{\textbf{Excitation and recovery}}\\
\midrule
Pulse & $P_{\rm exc}^{(0)}$ & $P_{\rm exc}^{(1)}/P_{\rm exc}^{(0)}$
& \shortstack{Bulk-pair\\share}
& \shortstack{Peak leakage\\$/P_{\rm exc}^{(0)}$}
& \shortstack{$\tret$ (ns)\\$p=0,\,1$}\\
\midrule
0.15 meV & $4.2\times10^{-2}$ & 1.05 & $<10^{-4}$ & 0.52 & 122, 131\\
0.20 meV & $1.9\times10^{-2}$ & 0.99 & $<10^{-4}$ & 0.54 & 130, 138\\
0.30 meV & $5.8\times10^{-3}$ & 0.99 & 0.69 & 0.59 & 131, 139\\
0.50 meV & $9.8\times10^{-5}$ & 1.01 & 0.94 & 0.70 & 129, 137\\
0.80 meV & $4.8\times10^{-9}$ & 1.00 & 0.95 & 0.76 & 126, 134\\
Broadband & $1.9\times10^{-4}$ & 1.00 & 0.36 & 0.59 & 128, 136\\
Thermal & $1.2\times10^{-2}$ & 1.10 & 0.003 & 0.52 & 117, 125\\
\bottomrule
\end{tabular*}

\vspace{7pt}
\begin{tabular*}{\textwidth}{@{\extracolsep{\fill}}lrrrr@{}}
\toprule
\multicolumn{5}{@{}l}{\textbf{Factorization error and coherence loss}}\\
\midrule
Pulse & \shortstack{Factorized residual\\$f_0/P_{\rm exc}^{(0)}$}
& \shortstack{Peak leakage\\factorized/exact}
& $(1-\chi)/\Pbar$
& \shortstack{Bulk-pair paths\\via Majorana level}\\
\midrule
0.15 meV & 1.86 & 0.98 & 1.00 & --\\
0.20 meV & 1.94 & 1.00 & 1.00 & --\\
0.30 meV & 1.98 & 1.69 & 0.75 & 0.64\\
0.50 meV & 2.00 & 1.66 & 0.81 & 0.80\\
0.80 meV & 2.00 & 1.65 & 0.82 & 0.82\\
Broadband & 2.00 & 1.32 & 0.90 & 0.73\\
Thermal & 1.96 & 0.99 & 1.00 & --\\
\bottomrule
\end{tabular*}

\vspace{7pt}
\begin{tabular*}{\textwidth}{@{\extracolsep{\fill}}llr@{}}
\toprule
\multicolumn{3}{@{}l}{\textbf{Numerical checks}}\\
\midrule
Quantity & Condition & Value\\
\midrule
$\Pflip(10~\mathrm{ms})$ & All pulses; $p=0,1$ & $<4\times10^{-35}$\\
Discarded probability & All pulses; $p=0,1$; $k_{\max}=2$ & $\le3.6\times10^{-6}$\\
Peak leakage & 0.20 meV & $1.06\times10^{-2}$\\
Time of peak leakage & 0.20 meV & 176 ns\\
Time to $\Pleak<10^{-6}$ & 0.20 meV & $1.4~\mu$s\\
Factorized $f_0(10~\mathrm{ms})$ & 0.20 meV & $3.76\times10^{-2}$\\
\bottomrule
\end{tabular*}

\vspace{4pt}
\begin{minipage}{\textwidth}
\footnotesize
$p=0,1$ labels the initial Majorana occupation.
Bulk-pair share: fraction of excitations creating two bulk quasiparticles.
The last column gives the fraction of bulk-pair excursions involving the Majorana level.
Residual occupations are evaluated at 10 ms; coherence losses are per wire.
\end{minipage}
\end{table}

\clearpage
\begin{table}[!htbp]
\centering
\small
\caption{Coherence (Fig.~\ref{fig:coherence}). $D=1-|\chi|$ and $D_{\rm ref}=D(\eps=1~\mu\mathrm{eV})$. Ratios $D/D_{\rm ref}$ are evaluated well below the crossover. Losses are per wire unless noted.}
\label{tab:coh}
\begin{tabular}{@{}p{0.29\textwidth}p{0.36\textwidth}p{0.29\textwidth}@{}}
\toprule
Quantity & Condition & Value \\
\midrule
\multicolumn{3}{@{}l}{\textbf{Uniform wire}} \\
Coherence loss & $\Omega_c=0.20$~meV & $1.926\times10^{-2}=\Pbar$ \\
$\hbar/(2\eps)$ & Working point & 0.44~ns \\
Pair $(1,2)$ decay rates & Capture (empty Majorana level); recombination (full); direct & $(3.0;\,2.9;\,2.4)\times10^7$~s$^{-1}$ \\
Final loss & $\Omega_c=0.50$~meV & $0.81\Pbar$ \\
Time to within 1\% of final loss & $\Omega_c=0.50$~meV & 250~ns \\
Time to leakage $<1\%$ of peak & $\Omega_c=0.50$~meV & $\approx780$~ns \\
Two-wire loss & Independent events & $(1.5$--$2.0)\Pbar$ \\
Broadband loss & $\Gesc=0;\,10^7$~s$^{-1}$ & $(1.76;\,1.73)\times10^{-4}$ \\
$\kappa_a$ & All retained modes & $<4\times10^{-10}$ \\
Relative change in $1-|\chi|$ & $\eps=0.001$~neV--1~$\mu$eV & $<3\times10^{-4}$ \\
\addlinespace
\multicolumn{3}{@{}l}{\textbf{Gate gradient}} \\
Lowest-mode centroid & $\delta U=0.4$~meV & $0.68L$ \\
$\kappa_1$ & $\delta U=0.2;\,0.4$~meV & 0.48; 0.89 \\
Loss at $\eps=1~\mu$eV & $\Omega_c=0.15$~meV, $\delta U=0.4$~meV & $6.6\times10^{-2}$ \\
Crossover $\hbar\Gamma_1/2$ & $\Omega_c=0.15$~meV, $\delta U=0.2;\,0.4$~meV & 9.0; 13.9~neV \\
All crossovers & Tested pulses and gradients & 2--45~neV \\
$D/D_{\rm ref}$ & $\Omega_c=0.15$~meV, $\delta U=0.4;\,0.2$~meV & 0.48; 0.86 \\
$D/D_{\rm ref}$ & $\Omega_c=0.20$~meV, $\delta U=0.4;\,0.2$~meV & 0.83; 0.92 \\
\bottomrule
\end{tabular}
\end{table}

\begin{table}[!htbp]
\centering
\small
\caption{Hybrid-state normalization (Fig.~\ref{fig:parent}). Rate ratios compare exact states with the effective model and bare vertex at $\gamma_t=\Delta_S$.}
\label{tab:hybrid}
\begin{tabular}{@{}p{0.29\textwidth}p{0.36\textwidth}p{0.29\textwidth}@{}}
\toprule
Quantity & Condition & Value \\
\midrule
$|Z_0-Z(0)|$ & & $\le1.2\times10^{-5}$ \\
Excess of $Z_a$ above $Z(0)$ & Low-energy states & $\le2.3\%$ \\
$Z_a$ & $E_a/\Delta_S=0.96$--0.97; $\gamma_t/\Delta_S=0.5;\,1;\,2$ & 0.072; 0.038; 0.025 \\
$H_{\rm eff}$ energy error & $E_a<0.43\Delta_S$ & $<5\%$ \\
22nd energy / $\Delta_S$ & Effective; exact & 1.63; 0.97 \\
Kernel ratio & Near threshold & $1/4$ \\
Cumulative-rate ratio & $\Omega>0.6\Delta_S$ & 0.22--0.27 \\
$Z_0Z_a$ & Threshold to $\Delta_S$ & 0.25 to 0.02 \\
Rate ratio; $Z_0Z_1$ & First excitation & 0.229; 0.251 \\
Rate ratios & Higher states paired by index; last state & 0.03--0.97; 2.9 \\
\bottomrule
\end{tabular}
\end{table}

\clearpage
\begin{table}[!htbp]
\centering
\small
\caption{LO wire and matched Kitaev chains (Fig.~\ref{fig:lom}). $K_\ell$ and $K_h$ denote low- and high-density chains; triples follow $(K_\ell,K_h,\mathrm{LO})$. Ranges cover the seven reference pulses. Efficiency ratios are chain / wire.}
\label{tab:lom}
\begin{tabular}{@{}p{0.29\textwidth}p{0.36\textwidth}p{0.29\textwidth}@{}}
\toprule
Quantity & Condition & Value \\
\midrule
\multicolumn{3}{@{}l}{\textbf{Recovery and coherence}} \\
$\Pflip$ & $t=10$~ms, all models & $<2\times10^{-34}$ \\
$\tret$ & $(K_\ell,K_h,\mathrm{LO})$ & 97--105; 135--158; 117--131~ns \\
Factorized $f_0/\Pexc$ & $(K_\ell,K_h,\mathrm{LO})$ & 1.81--2.00; 1.90--2.00; 1.86--2.00 \\
$(1-\chi)/\Pbar$ & $(K_\ell,K_h,\mathrm{LO})$ & 0.83--1.00; 0.70--1.00; 0.75--1.00 \\
$\kappa_a$ & All models & $<4\times10^{-10}$ \\
\addlinespace
\multicolumn{3}{@{}l}{\textbf{Excitation efficiency}} \\
$\eta_{K_\ell}/\eta_{\rm LO}$ & Narrowband $<0.5$~meV; thermal; broadband & 0.66--1.49; 1.34; 1.20 \\
$\eta_{K_h}/\eta_{\rm LO}$ & Narrowband $<0.5;\,0.8;\,1.0$~meV & 0.37--0.99; $7\times10^2$; $5\times10^4$ \\
Change in $\eta_{K_h}$ & $k_0a=1.0$ or 1.45; thermal, broadband & $<1\%$ \\
Change in $\eta_{K_h}$ & Same $k_0a$; narrowband $\le0.5;\,1$~meV & $<11\%$; $<35\%$ \\
\addlinespace
\multicolumn{3}{@{}l}{\textbf{Momentum and boundaries}} \\
Majorana-plus-bulk share & Coupling below 0.6~meV & $\approx2/3$ \\
$|k_a\pm k_0|$; Fourier tail & LO wire & $<0.05$~nm$^{-1}$; $q_x^{-6}$ or steeper \\
$k_a+k_0$; Fourier tail & $K_h$ & $\approx1$~nm$^{-1}$; $q_x^{-2}$ \\
Fourier-tail reduction & 100-nm edge window on transition densities; $|q_x|=0.1$--0.5~nm$^{-1}$ & $10^2$--$10^6$ \\
Bulk-pair share & $K_h$, near 1~meV & $\approx1/2$ \\
Coupling from $\theta<30^\circ$ & $\Omega=0.6$--1.2~meV; $K_h$; LO & 95\%; 0.4\% \\
Coupling from $\theta>70^\circ$ & $\Omega<0.6$~meV, all models & 79\%--95\% \\
Efficiency drop to 1~meV & From threshold; low; high density & $10^9$; $10^4$ \\
\addlinespace
\multicolumn{3}{@{}l}{\textbf{Field and chemical-potential scans}} \\
Gap minimum near $k_F$ & Field scan ($\mu=0$); gate scan & $V_Z>0.338$~meV; $\mu<0.234$~meV \\
Deviation of $\eta_{K_\ell}$ & In this region; thermal; broadband & $\le38\%$; $\le47\%$ \\
Narrowband deviation & Same region, 0.15, 0.20, 0.40~meV & Largest factor 2.2; mean excess 25\% \\
Thermal $\eta_{\rm LO}$ & $\mu=0.20;\,-0.30$~meV & 0.048; 5.3 \\
$\hbar v_F$ & $V_Z=0.225$~meV; $K_\ell$; LO $k_F$; LO $k=0$ & 4.8; 34; $\approx20$~meV\,nm \\
Modes below the $k_F$ minimum & $V_Z=0.225$~meV; $K_\ell$; LO & 21; 6 \\
Thermal $\eta_{K_\ell}/\eta_{\rm LO}$ & $V_Z=0.225$~meV; $\mu=0.375$~meV & 1800; 23 \\
\addlinespace
\multicolumn{3}{@{}l}{\textbf{Selection rule}} \\
$\langle\sigma_x\tau_x\rangle$ & Majorana spinors, $V_Z=0.225$~meV & $-0.89$ \\
Weak-line suppression (median) & $V_Z=0.225$--0.25~meV; versus $k_F$ lines within 0.1~meV above the minimum & 7--10 fold \\
Weak-line suppression (median) & Same range; versus $K_\ell$ lines & 18--28 fold \\
Thermal $\eta_{\rm LO}$ & Working point to $V_Z=0.225$~meV & 0.11--0.20 \\
Weak-line share & $V_Z=0.225$~meV & 60\% \\
$E_g$; fraction of reference gap & $V_Z=0.225$ or $\mu=0.375$~meV & 25~$\mu$eV; $1/5$ \\
Low-energy rates $\ge k_F$-branch rates & Chemical-potential scan & $\mu\ge0.35$~meV \\
Thermal $\eta_{\rm LO}$; enhancement & $\mu=0.375$~meV & 3.7; 31 \\
\bottomrule
\end{tabular}
\par\smallskip
\raggedright\footnotesize Weak lines lie below the $k_F$-branch minimum. Efficiencies are in \effu.
\end{table}

\clearpage
\end{document}